\documentclass[preprint,12pt]{elsarticle}

\usepackage{amssymb}
\usepackage{url}
\usepackage{booktabs} 
\usepackage{hyperref}
\usepackage{caption}
\usepackage{rotating}
\usepackage{graphicx}
\usepackage{multirow}
\usepackage{array}
\usepackage[table]{xcolor}
\definecolor{ug_color}{HTML}{AEA198}
\definecolor{imp_color}{HTML}{B1B3C2}
\definecolor{dom_color}{HTML}{9F8CA7}
\definecolor{data_color}{HTML}{79A099}
\definecolor{ux_color}{HTML}{B8AB77}
\definecolor{tr_color}{HTML}{7695AA}
\definecolor{int_color}{HTML}{958886}
\definecolor{mod_color}{HTML}{B0B1B1}
\definecolor{ag_color}{HTML}{958886}
\definecolor{md_color}{HTML}{80A1A1}   

\begin{document}

\begin{frontmatter}


% \author{Name\corref{cor1}\fnref{label2}}

% \author{Anonymous Author(s)}
% \author[1]{Muhammad Raees\corref{cor1}\fnref{label2}}
\author[1]{Muhammad Raees}

% \cortext[cor1]{Muhammad Raees (mr2714@rit.edu)}
% \fntext[label2]{Equal Contribution}
% \cormark[1]
\ead{mr2714@rit.edu}
\affiliation[1]{organization={Rochester Institute of Technology},
            city={Rochester},
%          citysep={}, % Uncomment if no comma needed between city and postcode
            state={NY},
            country={USA}}
\author[2]{Inge Meijerink}
% \author[2]{Inge Meijerink\fnref{label2}}
% \cortext[cor1]{Muhammad Raees (mr2714@rit.edu)}
% \cormark[1]
\ead{ingeemeijerink@gmail.com}
\affiliation[2]{organization={Utrecht University},
            city={Utrecht},
%          citysep={}, % Uncomment if no comma needed between city and postcode
            % state={NY},
            country={The Netherlands}}

\author[2]{Ioanna Lykourentzou}
% \cortext[cor1]{Muhammad Raees (mr2714@rit.edu)}
% \cormark[1]
\ead{i.lykourentzou@uu.nl}

\author[3]{Vassilis-Javed Khan}
% \cortext[cor1]{Muhammad Raees (mr2714@rit.edu)}
% \cormark[1]
\ead{javed.khan@sappi.com}
\affiliation[3]{organization={Sappi Europe},
            % city={Utrecht},
%          citysep={}, % Uncomment if no comma needed between city and postcode
            % state={NY},
            country={Belgium}}

\author[1]{Konstantinos Papangelis}
% \cortext[cor1]{Muhammad Raees (mr2714@rit.edu)}
% \cormark[1]
\ead{kxpigm@rit.edu}

\title{From Explainable to Interactive AI: A Literature Review on Current Trends in Human-AI Interaction}


% \author[inst1]{Author One}

% \affiliation[inst1]{organization={Department One},%Department and Organization
%             addressline={Address One}, 
%             city={City One},
%             postcode={00000}, 
%             state={State One},
%             country={Country One}}

% \author[inst2]{Author Two}
% \author[inst1,inst2]{Author Three}

% \affiliation[inst2]{organization={Department Two},%Department and Organization
%             addressline={Address Two}, 
%             city={City Two},
%             postcode={22222}, 
%             state={State Two},
%             country={Country Two}}

\begin{abstract}
AI systems are increasingly being adopted across various domains and application areas. 
With this surge, there is a growing research focus and societal concern for actively involving humans in developing, operating, and adopting these systems. 
Despite this concern, most existing literature on AI and Human-Computer Interaction (HCI) primarily focuses on explaining how AI systems operate and, at times, allowing users to contest AI decisions.
Existing studies often overlook more impactful forms of user interaction with AI systems, such as giving users agency beyond contestability and enabling them to adapt and even co-design the AI's internal mechanics. 
In this survey, we aim to bridge this gap by reviewing the state-of-the-art in Human-Centered AI literature, the domain where AI and HCI studies converge, extending past Explainable and Contestable AI, delving into the Interactive AI and beyond. 
Our analysis contributes to shaping the trajectory of future Interactive AI design and advocates for a more user-centric approach that provides users with greater agency, fostering not only their understanding of AI's workings but also their active engagement in its development and evolution.
\end{abstract}


% \begin{highlights}
% \item Research highlights 1
% \item Research highlights 2
% \end{highlights}


\end{frontmatter}

% \linenumbers

\section{Introduction}
Artificial Intelligence (AI) is being incorporated in almost all aspects of professional and personal spheres, including healthcare, education, leisure, and business \cite{belic2019artificial, poola2017artificial, raaijmakers2019artificial, roll2016evolution}, partially owing to the growing accessibility of computing resources and the abundance of available data \cite{fradkov2020early}. 
Modern AI tools (e.g., ChatGPT \cite{chatgpt}, DALLE-3 \cite{dalle}, Stable Diffusion \cite{stablediffusion}) shape public opinion towards the benefits, rather than the drawbacks, of these technologies, by being accessible and allowing not only experts but also end-users to experiment directly with their capabilities. 
At the same time, the reliance on AI systems for automatic and autonomous decision-making raises critical concerns on issues such as copyrights, plagiarism, misconduct, and the spread of fake information \cite{beatrice2023online, zhong2023copyright}, as well as the repercussion of AI’s incorrect decision-making on human lives, including agency loss, privacy loss, bias, and discrimination \cite{mccormack2020design, Xiang2023online, collier2023online}. 
AI is still technology-centered---with its efficiency largely being measured based on system performance metrics rather than the quality of its interaction with humans, especially in practical applications \cite{xu2019toward}.
On the research front, recent studies have explored methods to make AI systems more adaptable by users and to gain their trust, examining human satisfaction, traceability, or collaboration \cite{honeycutt2020soliciting, dzindolet2003role, fugener2021will, puntoni2021consumers}. 
Efforts are also underway for contestability---the ability to oppose and contest AI decision-making \cite{alfrink2020contestable}. 
However, the application of these methods in practice is still limited.

When it comes to literature evaluation aimed at identifying trends and gaps in future human-AI interactions, the majority of research is concentrated on the explainable AI---to make systems understandable by users \cite{vereschak2021evaluate, dwivedi2023explainable, dovsilovic2018explainable}.
While this marks progress compared to early studies that solely assessed AI by (model) performance, these studies adopt the lens of a rather \textit{``passive''} human role in decision-making, which typically stops at the level of receiving explanations for AI decisions or, occasionally, at the ability to contest them. 
Beyond that, current implementations provide limited agency to users to control and adapt AI systems to their needs.
Active human-AI interaction involves continuous collaboration between users and AI systems, ultimately advocating for more human-centered (a.k.a. true \textit{``human-in-the-loop''}) approaches for system design \cite{xu2019toward, shneiderman2020human, xu2023transitioning}. 
However, having an overarching human agency with AI systems also becomes challenging to handle without adhering to human-centered approaches. 
For instance, the explainability of a system may interfere with its performance/complexity, and having more interactions may make the system susceptible to user fallibility. 
Therefore, a balanced human-AI interaction empowers users to take actions, separates user and AI tasks, and establishes a shared responsibility between the user and the AI system \cite{shneiderman2020human}. 
Research has emphasized the need for human-centered approaches from different perspectives such as transparency, trust, or interaction guidelines \cite{xu2019toward, shneiderman2020human, amershi2019guidelines, vereschak2021evaluate, li2021survey, mohseni2021multidisciplinary, amershi2014power}.
However, the majority of explorations still primarily focus on explanations as the main assessment criteria of human-AI approaches/interactions.
To the best of our knowledge, no systematic literature review study has examined contemporary developments in user interaction that are focused on fostering a more active role for the user in their interaction with the AI system beyond explanations.

\textbf{Our literature review aims to address this gap by consolidating and critically analyzing complementary state-of-the-art approaches in human-AI interaction, including studies from the Human-Centered AI, Explainable AI, and Interactive AI fields, among others, which extend beyond explanations and involve \textit{``explicit''}, \textit{``intentional''}, and \textit{``informed''} user interactions with AI systems.}  

%Results overview
Our systematic analysis, guided by the above-stated research agenda, identifies key research trajectories, patterns, and gaps, focusing on the active user interaction with AI systems. 
Among our key findings, we uncover that a wide range of studies fail to include end-users in the co-design of AI systems or even in simpler forms of interaction with them. 
In terms of applications, we find that recent research predominantly explores active interaction in low-risk areas such as education, leisure, and sports, evaluating trivial tasks, while neglecting high-risk domains such as healthcare and security. 
Regarding the goals of AI systems in their interactions with users, we observe that most systems concentrate on enhancing user experience to ultimately support user acceptance. 
Our analysis also reveals that only a handful of studies permit active modification of AI mechanics, particularly in the Interactive Machine Learning (IML) domain. 
IML explorations present an opportunity for designing systems that provide a higher degree of agency and user control, extending beyond system understandability. Finally, across all identified dimensions of analysis (users, implementations, goals), spanning different fields per dimension, we note that a significant portion of related literature proposes theoretical rather than practical solutions. 
This suggests an opportunity for the growth of Interactive AI towards more practical experimentation to understand and shape the emerging field of active AI interaction.

%Contributions
Overall, the contributions of this work are summarized as follows:
\begin{itemize}
    \item We present a systematic literature analysis of the current state and bottlenecks in research and practice regarding explicit, intentional, and informed human-AI interaction. Our analysis incorporates contributions from various fields, including Human-Centered AI, Collaborative AI, Contestable AI, Interactive AI, Interactive Machine Learning, and Hybrid Intelligence.  
    \item We critically evaluate relevant literature, from the perspective of users, implementations, and goals of AI systems. We then examine practical interconnections, providing an overview of AI's impact on user trust, acceptance, and utilization.
    \item We synthesize challenges and research gaps, focusing on balancing system autonomy, user agency, and user needs across various aspects, such as user experience, transparency, interaction, and augmentation. 
    \item We provide guidelines for future system design to overcome the identified challenges and gaps toward a more efficient and balanced interaction of humans with AI systems and vice versa. These guidelines advocate for participatory human-AI interaction design and, for reflective evaluation by the relevant stakeholders.
\end{itemize}

The rest of this paper is structured as follows. Section \ref{section:2back} presents a brief background of this work, providing an overview of the fields of explainable AI and human-AI interaction. 
Section \ref{section:3method} explains our methodology, covering the inclusion/exclusion criteria, search, data collection, study evaluation protocols, and definitions/terminologies. 
Section \ref{section:4analysis} delves into our analysis across three main dimensions: 1) AI Users, 2) AI Implementations, and 3) AI Goals, with each discussing further sub-dimensions. 
We accompany the analysis of each of those dimensions with a summary of our findings to help the reader easily grasp observed nuances in the state-of-the-art. 
Section \ref{section:5discuss} discusses our findings synthetically, including the extent to which current literature fulfills the challenges and needs for achieving user-centered interactivity in practice, as well as two common perspectives of interaction discussed in the literature, namely collaboration and augmentation. 
Furthermore, we discuss the critical issue of agency in human-AI interaction research, highlighting a nascent but mostly unexplored trend for user inclusion in the development and evolution of AI systems. 
Finally, we discuss issues related to interfacing and outline what is needed for current research to effectively address interaction challenges. 
We also discuss the limitations and suggest future directions for this work. 
Section \ref{section:6conclude} concludes with the key findings of this work.

\section{Background}
\label{section:2back}
AI witnessed strong performance growth in the past few decades, transitioning from the phase of mimicking human intelligence to algorithmic dominance \cite{mccarthy2006proposal, buchanan2005very, epstein2015wanted}.
Human imitation, such as learning, adapting, predicting, and interacting, has remained fundamental to AI \cite{annoni2018artificial, dotov2020dynamic}. 
However, with the growth in data-intensive computing, newer techniques such as machine learning (deep neural networks) and learning everything directly from data emphasize predictive accuracy over the human agency \cite{copeland2000modern, fradkov2020early}.
Such developments emphasize measuring the success of a system by its superior predictive accuracy over others.
For instance, many research studies state the supremacy of their method by outperforming others in contention, piling up on complexity, and focusing less on human influence, factors, or needs \cite{xu2023transitioning}.
However, with widespread AI applications in user domains, the accuracy-oriented metric is exposed to challenges of AI adoption and acceptance in various stakeholder contexts, for instance, in cases where users need to understand or contest the decision made by the system.  

Growing demand for intelligent systems (those fed on huge data or computation) in numerous domains leads to complex AI systems (such as deep neural networks with billions of parameters).
Complex AI systems are referred to as \textit{``black boxes''} that obfuscate the transparency of their decisions, concealing their internal working from the user \cite{guidotti2018survey}.
Tracing how a system learns and makes decisions becomes challenging with rising complexity \cite{arrieta2020explainable}.
Oftentimes, it is difficult to trace how a gigantic deep neural network works, inhibiting the overall understanding of how it makes decisions. 
Consequently, such systems have raised many ethical questions, making it difficult to apply algorithmic decisions to the high-stake fields in real-world scenarios, where tracing is imperative \cite{guidotti2018survey}. 
For example, this could happen in various contestable AI applications such as medicine, law, finance, security, etc., where the rationale for decisions made by the system is important for end-users. 
Traceability and contestability approaches try to curb a lack of understanding about decisions for their users. 
For instance, in medical domains, experts often make decisions with causal inference, thus requiring far more information to support their diagnosis \cite{tjoa2020survey, plass2022understanding}.
As described in Plass et al. \cite{plass2022understanding}, understanding and rationalizing decision-making in medical diagnosis \cite{holzinger2021toward} necessitates utilizing additional information. For instance, a clinician with years of experience and knowledge considers several factors before diagnosing a medical case.
These factors may include initial assessments derived from patient data, medical history, and/or physical examinations. 
Incorporating such complementary information enables clinicians to causally support or refute the decision path for the diagnosis \cite{plass2022understanding, holzinger2021toward}. 
Consequently, the demand for methods that render AI systems more transparent and interpretable, thereby justifying their decisions with complementary or explanatory information, grows.

\subsection{Explainable AI (XAI)}
Explainable AI (XAI) \cite{xu2019explainable}, in a nutshell, helps users understand how an AI system makes a decision \cite{montavon2018methods}. 
XAI encounters opaque AI systems with explanations and traceability for the decisions made by those.
Explainability of AI systems has been explored in several dimensions, however, there are two widely accepted strategies across studies \cite{xu2019explainable, dwivedi2023explainable}. 
The bifurcation between these two strategies is obvious where one advocates using transparent/traceable models (i.e., direct interpretations, like rule-based models) while the other focuses on enhancing the interpretations of opaque models \cite{liao2021human}.
The former leads to simpler and more acceptable models, however, the performance of such models also degrades in complex settings.
Opaque models have complex learning architectures, leading to better performance compared to simpler models, yet their interpretation is always a concern \cite{liao2021human, arrieta2020explainable, dovsilovic2018explainable}.
Therefore, the latter strategy (i.e., Post-hoc explainability) aims to enhance explanations of such opaque AI models \cite{liao2021human}.
Such models often utilize output tracing, for instance, by input perturbation to observe how the models' outcomes change or to understand the internal working of models by their stakeholders \cite{koh2017understanding}. 
However, for an end-user, such explanations or traces are still very complicated to understand concretely.

XAI, being at the forefront of making complex AI systems more acceptable, has made substantial developments to augment the user understanding of how AI systems make decisions \cite{dwivedi2023explainable}.
Although there are countless calls for XAI, there is still a substantial lacuna in research for studying the efficacy of the users in understanding and interacting with AI systems \cite{dovsilovic2018explainable, abdul2018trends}. 
XAI has made many inroads in making AI systems more acceptable, however, with AI systems having increased real-world implications and user integration, XAI does not suffice every purpose.
In addition, evaluating XAI approaches with different usage contexts is still in its infancy with several attempts to make user involvement effective and adaptive, addressing human factors ranging from expertise to cognitive understanding \cite{suresh2021beyond, liao2022connecting, kim2023help, stefik2023roots}.
Evidence from several comprehensive studies \cite{dwivedi2023explainable, lai2023selective, liao2021human} corroborates that transparency and understanding of AI systems are fundamental to users. 
However, many of these studies only enhance the interpretation behind the decision-making of AI systems rather than focusing on how users interact with such explanations. 
Recent work highlights the significance of varying stakeholders' contexts, including for the end-users, in XAI application deployments \cite{dwivedi2023explainable, liao2021human}. 
Also, different stakeholders will have different goals and might have significantly different explainability needs altogether from the same system \cite{suresh2021beyond}. 
Stakeholder involvement in system design (e.g., debugging, evaluation, compliance, or experience) is important for effective usage \cite{liao2022connecting}. 
This involvement varies according to the stakeholders' needs for the system explanation.
Therefore, the explanation mechanism should be adaptive to usage contexts necessitating more active user involvement in the process. 
Studies have utilized established frameworks (interview studies, surveys, etc.) for eliciting users' XAI needs \cite{suresh2021beyond, liao2022connecting, kim2023help, lai2023selective}.
However, the current research focus still neglects to examine how these systems integrate into established processes thoroughly, and how these are adapted to varying contexts, yet offering limited opportunities for interaction to end-users
\cite{yampolskiy2019predicting}.

\subsection{Human-AI Beyond Explainability}
With the ever-increasing integration of AI into numerous domains, human involvement and interaction with AI systems increase proportionally. 
This necessitates an active user role with AI systems fostering better collaboration and acceptance \cite{shneiderman2020human, shneiderman2020human_rel}. 
Human-AI interaction is an area of development to bridge the gap between AI and humans, stepping on the intersection of AI and Human-Computer Interaction (HCI) \cite{xu2019toward}. 
The intersection, being highly elastic, is an umbrella covering the sole human involvement in the process of AI development (a.k.a. \textit{``humans-in-the-loop''}) to empowering end-users as co-creators of (\textit{``Interactive AI''}) systems. 
Enhancing the role of humans is a core principle in interactive AI, which is often mediated through some form of collaboration \cite{epstein2015wanted}. 
However, the existing focus on AI performance often overlooks the collaboration with users \cite{renz2021reinvigorating}.
Utopian collaboration, although challenging to achieve, strengthens the expertise of both humans and AI and overcomes their limitations \cite{shneiderman2020human}. 
For instance, users complement the AI where it falls short and vice versa. 
However, several factors (such as human expertise, fallibility, and socio-technical issues) inhibit the ideal collaboration \cite{heer2019agency, xu2019toward}.
Therefore, most collaborations, generally, restrict humans as data providers and feedback agents to merely act as consumers of AI systems \cite{newlands2021lifting, chignell2021human, allen1994mixed}. 

Improper integration and collaboration may also undermine the benefits of human-AI interaction \cite{strauch2017ironies, honeycutt2020soliciting}, necessitating the development of a well-balanced interaction framework.
Nevertheless, the inclusion of humans in the loop has shown promising results toward improving decision-making, trust, and acceptance of AI systems \cite{van2021hybrid, agrawal2022power}.
Human-AI collaboration is always influenced by factors such as user expertise, their trust in the system, and the ability to have control over it \cite{inkpen2023advancing}, for instance when utilizing it to assist decision-making \cite{kuang2023collaboration}.
Interaction can support assisting users rather than replacing them \cite{annoni2018artificial, dotov2020dynamic}.
Hence, instead of making AI better than humans, goals translate to expanding and supporting human creativity \cite{mccormack2020design}.
To conceptualize interactivity, the essence of the human-centered approach positions humans not merely as participants in the loop but places them at its very core, for instance, by constructing AI systems with a central focus on users, and inverting the conventional paradigm of humans adapting to the technology \cite{shneiderman2020human}.
This narrative is incontrovertible with the widespread penetration of AI systems within human lives, calling for approaches and guidelines for AI development and making humans the center of the design process \cite{auernhammer2020human}. 
Pioneering work with similar aspirations by Amershi et al. \cite{amershi2014power} and Xu et al. \cite{xu2023transitioning} highlight the need for empowering end-users in AI systems.
Big technology companies such as Google, Microsoft, and IBM \cite{googleaiguidebook, amershi2019guidelines, ibmdesignforai} have also defined guidelines for building AI systems from a user-centered perspective. 
However, to what extent the current human-AI research and development approaches can cater to this shift is still under-explored.
Hence, more interactive explorations are needed to better understand human participation in AI decision-making.

\textbf{\textit{Summary.}} A critical research focus lies in identifying methods for explanations and improving the adoption of opaque AI systems with different stakeholders. 
We highlighted the limitations of user interaction and current issues with the opaqueness (limited interpretation) of AI systems. 
XAI has been the central pillar in making AI more transparent and interpretable for users.
XAI has made strides towards achieving this goal to an extent, many domains have far more integration of expert and non-expert users with the AI systems, bringing countless human factors into play.
XAI primarily deals with the passive role of humans in the loop merely going beyond explanations.
Hence, the problem transcends beyond explainability, from contesting decisions to adaptations, requiring effective human-AI interaction.
The human-AI interaction, beyond explainability, advocates for agency and an active role in the co-creation of AI systems.
Hence, it is critical to examine to what extent the state-of-the-art has progressed in this aspect. 
Whether current HCI practices sufficiently support developing interactive human-AI systems and whether other practical interpretations are being explored.
Hence, a systematic literature evaluation \cite{mulrow1994systematic}, from the lens of human-AI interaction is essential to establish state-of-the-art for current practices that foster user agency and interactive adaptation/control of AI systems.

\section{Methodology}
\label{section:3method}
In this section, we elaborate on the scope and evaluation criteria of our literature review. 
We use a method-based developmental literature evaluation \cite{palmatier2018review, templier2015framework} to investigate the interactivity in AI systems. 
Fundamental to systematic reviews, we utilize protocols for the assessment of the research studies  \cite{keele2007guidelines}. 
Protocols, being highly encouraged in literature research, enhance the quality of the evaluation process (e.g., to document the analysis and ensure consistency). 
Preferred Reporting Items for Systematic Reviews and Meta-Analyses (PRISMA) \cite{PAGE2021178} is an effective methodological protocol for document analysis and assessment.
Therefore, with the PRISMA, we followed a systematic and structured procedure to collect and evaluate research papers for our study. 
With initial evaluation, we identified the relevant keywords for search engine querying. 
Then, based on those keywords, we collected a data set of primary studies \cite{meade1997selecting, bartels2013perform} and followed a snowballing \cite{wohlin2014guidelines} procedure to enhance the data set. 
The following sections explain the processes of protocol execution. 

\subsection{Inclusion and Exclusion Criteria}
Before selecting our keywords, we ensured to collect as many relevant articles as possible by formulating very open inclusion and exclusion criteria encompassing studies in the human-AI interaction domain.
As interaction was the core focus of this research, we did not focus on purely technical or algorithmic AI studies.
Instead, our focus was to identify studies openly featuring interaction between humans and machines and analyze how those interactions are perceived. 
Studies dealing with machine-to-machine interactions were excluded, as our context remains only on the interactions between machines and humans.

At a later stage, we enforced stricter inclusion and exclusion criteria, only considering human-AI interaction studies if the interaction with the user is \textbf{explicit}, \textbf{intentional}, and \textbf{informed}.
By \textbf{explicit} we mean that the interaction is obvious to the user and it will impact system functions.
By \textbf{intentional} we mean the user initiates the interaction (i.e., by exercising agency) and remains in control of it, including the notion that the user also understands the impact of their interaction and the functionality of the system.
By \textbf{informed} we mean that the user is aware of the interaction and does not simply act passively as a data provider in the loop.
In the scope of our study, a system is not considered to have an effective interaction if it lacks any one of these three aspects.
For instance, if the system considers users passively to infer their intent (e.g., inferring their interest in a given product based on user activity) without explicit interaction and dialogue with the user, the interaction is not considered to be explicit, intentional, and informed.  
In another example, if a system does not allow the user to understand its functionality, either intuitively or through explanations, the interaction is not explicit or informed. 
Studies that only provide the user with explanations without affording them the possibility of further interaction are also put out of scope, since, in this case, the interaction is informed, but neither explicit nor intentional.

To gather the literature on the subject, we performed an explanatory search to select keywords from renowned libraries such as ACM, Springer, Elsevier, and Science Direct.
By screening venues closer to human-AI interaction, we identified and finalized our search strings to query the search engines.
The following popular search terms are included in our representative search strategy. 

\begin{itemize}
    \item Explainable Artificial Intelligence, Explainable AI, XAI
    \item Contestable Artificial Intelligence, Contestable AI, ContestAI
    \item Collaborative Artificial Intelligence, Collaborative AI, CollabAI
    \item Hybrid Intelligence, Hybrid AI
    \item Interactive Artificial Intelligence, Interactive AI
    \item Human-centered Artificial Intelligence, Human-centered AI, HCAI
    \item Interactive Machine Learning, Interactive ML, IML
\end{itemize}

Each search term identifies a different type of interactivity with the user.
Well-accomplished abbreviations such as XAI, AI, or ML were also embedded where necessary.
XAI, with a strong recent interest, grasps the current mainstream level of explanations for users. 
Contestable AI allows the user to object to a decision made by the system.
Collaborative AI focuses on integrating tasks between humans and AI by forming teams.
Hybrid AI is more of a mixed interaction and is close to collaborative AI. 
The next three terms scout for the level of interactivity we truly want.
Interactive AI/ML was expected to help us retrieve work with a similar focus. 
Finally, human-centered AI links to works purely on user-centered AI design. 

\subsection{Search and Data Collection}
For more unbiased results, we did not focus on any specific library to search for our desired results.
Based on search terms and the inclusion criteria, we utilized Google Scholar \cite{scholar} and Scopus \cite{scopus} to identify research material, as well as to ensure search diversity. 
Google Scholar, a widely used scientific search engine, offers various search and filter options to get unbiased and a wider range of articles.
We scanned titles from thousands of initial results to identify the primary collection.
With a focus on finding research papers, we excluded books, chapters, editorials, notes, erratum, and letters. 
We included articles, conference papers, reviews, and short surveys as they all include some level of detailed work and/or are peer-reviewed.
Also, we focused on recent studies and mainly identified/filtered (initial) research studies after 2003 (only 6 known studies before that).
Search results were then collected from renowned libraries such as ACM Digital Library, Springer Link, ScienceDirect, Taylor, and Francis, among other lesser-known sources to reduce bias towards specific venues.

\subsection{Study Eligibility Evaluation}
We commenced the review process with an initial extraction of 268 relevant studies.
We removed a total of 13 articles that were not written in English or were not retrievable.
After the initial identification and selection of studies, three authors performed an eligibility criteria assessment on the remaining 255 articles to choose representative studies, enforcing inclusion/exclusion criteria. 
The process followed a two-step approach: 1) evaluating studies based on title, venue, and abstract and 2) using full-text versions of the papers that were closer to selection in Step 1.
Initially, we used a Google Sheet to code studies into different levels to specify how the paper relates to human-AI interactions.
To ensure the trustworthiness and quality of the coding process, the third author then analyzed the coding of the other two authors and solved issues of conflicts on classification. 
With an intraclass correlation coefficient \cite{koo2016guideline} of 0.94, there was a high agreement rate and excellent reliability of the categorization between the authors. 
Pending issues were discussed collectively to reach a consensus on the final classification.
This phase also ensured that we only considered papers where human involvement is considered or anticipated.

\begin{figure}[h]
  \centering
  \includegraphics[width=\linewidth]{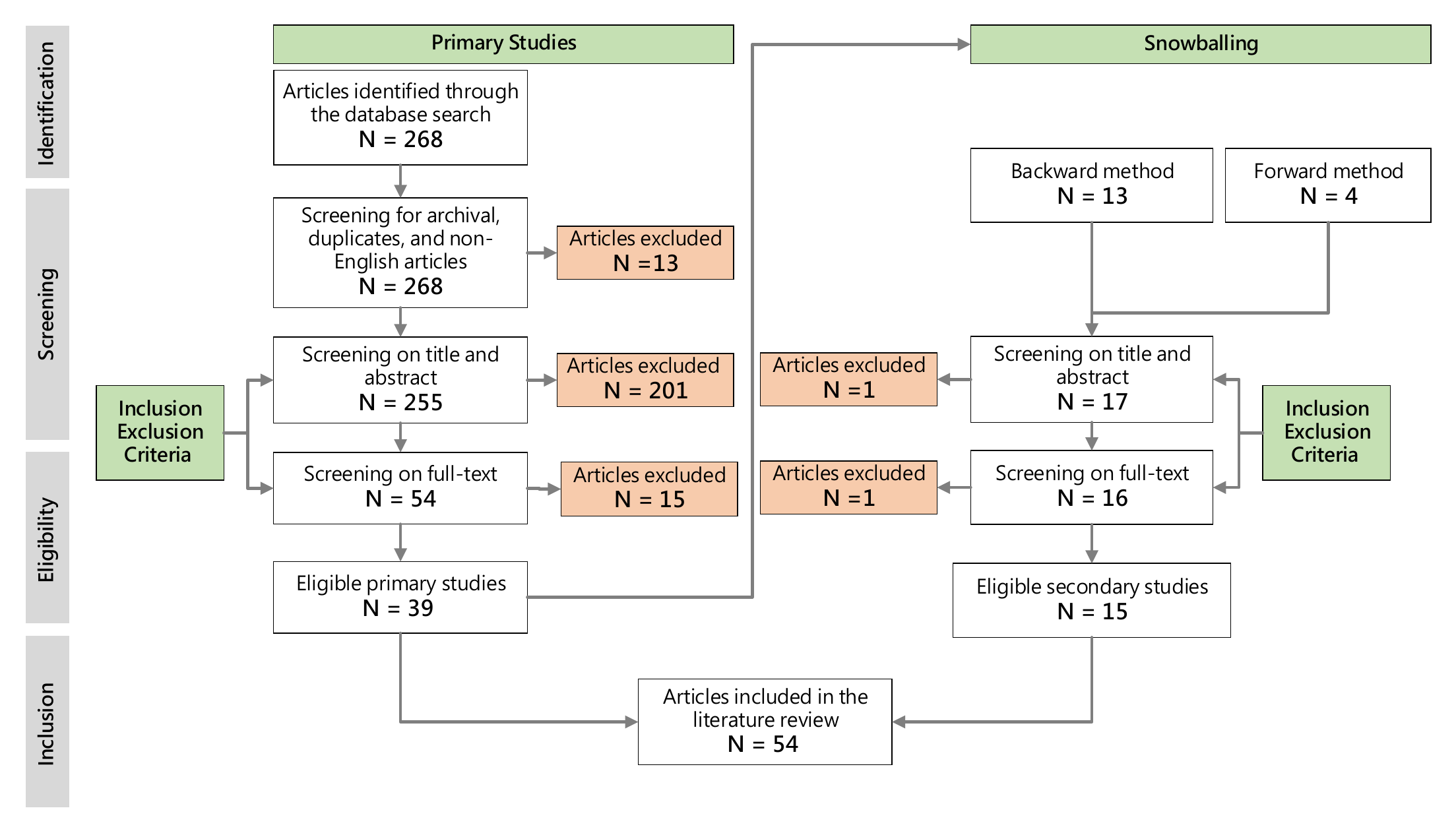}
  \caption{Outline of the step-by-step process followed to find, review, and assess studies. We adapted the procedure from PRISMA \cite{PAGE2021178}, a widely recognized set of guidelines for systematic reviews.}
  \label{fig:prisma}
  % \Description{Flow of the data collection and analysis process}
\end{figure}

\begin{table}
  \caption{Unique publications identified per search term and engine. After applying the inclusion and exclusion criteria, we retained 39 primary studies out of 268. From these, we retrieved an additional 15 unique studies using snowballing, resulting in 54 studies in total selected for the survey. Most of the retained studies are from the Human-centered and Collaborative AI fields, while all examined studies from the XAI were excluded as those only focus on explainability without further interaction of the user with the AI system.}
  \label{tab:searchterms}
  \scriptsize
  \begin{tabular}{|l|l|l|l|l|l|l|}
    \hline
    \multirow{2}{*}{Search Term} & \multicolumn{3}{c|}{Identified Studies} & \multicolumn{3}{c|}{Retained Studies} \\
    \cline{2-7}
    & Scholar \cite{scholar} & Scopus \cite{scopus} & Total & Primary & Snowballing & Total \\
    \hline
    XAI & 31 & 27 & 58 & 0 & 0 & 0 \\ \hline
    Contestable AI & 30 & 0 & 30 & 1 & 0 & 1 \\ \hline
    Collaborative AI & 30 & 16 & 46 & 8 & 7 & 15 \\ \hline
    Interactive AI & 30 & 20 & 50 & 6 & 0 & 6 \\ \hline
    Human-Centered AI & 30 & 20 & 50 & 7 & 3 & 10 \\ \hline
    Interactive Machine Learning & 23 & 0 & 23 & 13 & 3 & 16 \\ \hline
    Hybrid Intelligence & 11 & 0 & 11 & 4 & 2 & 6 \\ \hline
    Total & 185 & 83  & 268 & 39 & 15 & 54 \\ \hline
  \end{tabular}
\end{table} 

We employed five categories (levels 0 to 4), classifying studies into \textit{``Duplicate or unusable''}, \textit{``Purely (algorithmic)''}, \textit{``Including user testing''}, \textit{``Human-in-the-loop (as data provider)''}, and \textit{``User interaction''} based on their relevance to interactivity. 
Through this rigorous process, we eliminated most of the studies that did not meet the inclusion criteria (i.e., user interaction) and derived 39 primary studies for further analysis. 
The remaining articles lacked at least one of the inclusion criteria as interaction being explicit, informed, and intentional. 
The levels are explained as follows:

\begin{itemize}
    \item \textbf{Level 0:} Any remaining duplicates or being off-topic from interactions.
    \item \textbf{Level 1:} Purely algorithmic AI. Lack of user inclusion with a focus on the technical system aspects.
    \item  \textbf{Level 2:} User merely as a receiver of information, e.g., which method works best for the user without actively including the user.
    \item \textbf{Level 3:} User as a human-in-the-loop and (passive) data provider, e.g., human labor is used for data preparation tasks.
    \item \textbf{Level 4:} Active and explicit interaction by the user beyond merely providing data, e.g., providing feedback or modification to the system.
\end{itemize}

\begin{figure}[h]
  \centering
  \includegraphics[width=\linewidth]{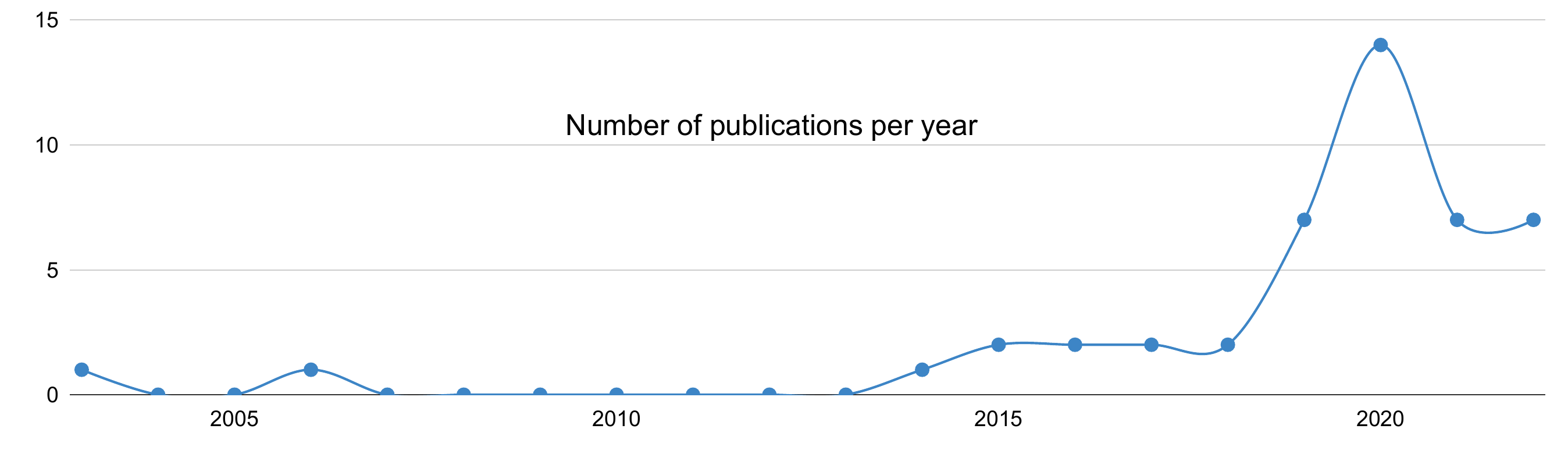}
  \caption{Number of included studies per year fulfilling inclusion criteria. The increase in human-AI publications in recent years is evident with a higher number of studies included in the survey that were published after 2018.}
  \label{fig:pubs_by_year}
  % \Description{Line chart showing how many studies are included per year.}
\end{figure}

For the snowballing procedure, we were only interested in the articles categorized as level 4. 
We completed a single iteration of both forward and backward snowballing.
We used Google Scholar to perform forward snowballing to identify papers that cited primary studies meeting the inclusion criteria. 
The snowballing ensured that scattered interactive AI literature was searched again with a different approach to reduce the risk of missing relevant work. 
Snowballing iterations were subjected to the same eligibility screening process, adding 15 studies to the existing 39. 
Figure \ref{fig:prisma} shows the overview of the protocol for eligibility evaluation, while table \ref{tab:searchterms} shows descriptive statistics of representative studies identified and retained for each term.
XAI and Contestable AI contribute little due to the lack of studies that directly measure interactivity with users (i.e., lacking explicit, informed, and intentional interactions).
Collaborative AI, Human-centered AI, and Interactive ML with 15, 10, and 16 articles, respectively, are the main focus of the study.
Interactive and Hybrid AI contribute equally, with 6 studies each.  
Table \ref{tab:publishers} shows the source description (i.e., libraries databases) of selected studies.
Figure \ref{fig:pubs_by_year} shows the trend of included studies over the years, with only two studies included in the survey before 2014. 
In almost all categories, around 75\% of the papers have been published in the last five years. 
The surge shows a growing interest in interactivity, the user experience, and understanding perspectives of AI systems for user contexts. 

\begin{table}
  \caption{Representation of libraries and publishers to the included studies in the survey. The majority of included studies are published and retrieved from ACM, Springer, and AAAI. The selection also shows diversity and inclusion from other libraries.}
  \label{tab:publishers}
  % \scriptsize
  \begin{tabular}{|l|c|c|}
    \hline
    Publication Venue & Number of Papers & Total Percentage \\ \hline
    ACM & 20 & 37.0\% \\ \hline
    SPRINGER & 6 & 11.1\% \\ \hline
    AAAI & 4 & 7.4\% \\ \hline
    TandF & 3 & 5.6\% \\ \hline
    ELSEVIER & 3 & 5.6\% \\ \hline
    IEEE Explore & 2 & 3.7\% \\ \hline
    Others & 16 & 29.6\% \\ \hline
    Total & 54 & 100.0\% \\ \hline
  \end{tabular}
\end{table}

\subsection{Definitions and Terminology}
Before analyzing the identified literature in detail, we review the existing definitions and terminologies around the referenced concepts in the human-AI literature \cite{weld2019challenge, guidotti2018survey, mohseni2021multidisciplinary}. 
Human-AI is defined as an integrated perspective of technology, people, and policies \cite{auernhammer2020human}.
The definition describes an intersection of HCI and AI, putting more focus on humans, such as through user-centered design that considers human factors rather than being technology-centered \cite{shneiderman2020human, xu2019toward}. 
To focus on interactivity, approaches must strive for user-centered aspects including agency, augmentation, and collaboration built into the design.
These terms are often used in conjunction with interactive AI/ML, or in general, with human-centered AI.
Explainability---as discussed in section \ref{section:2back}, also closely links with the human-centered approach.

\subsubsection{Interactivity and Agency}
An AI system should form a suitable form of interaction with the target stakeholders it directly affects.
User interactions with AI can be classified through the scale of interactivity they have with the system.
For instance, we need to ask how a user can interact with the AI system? what is the interaction method? or what is the purpose of interaction? to scale interactivity.  
Interactivity can take several forms, such as explicit or implicit, leading to consuming (passive) explanations, actively providing input/feedback on decisions, and contesting or correcting system decisions altogether (agency).
The interaction with AI allows users to achieve goals in various ways.
For example, some systems allow interaction for feedback \cite{ramos2020interactive} while others allow the agency to correct decisions \cite{teso2019explanatory}.
Some systems allow implicit interaction through gestures, natural language, or experience feedback \cite{patel20193d, nicholls2018collaborative}.
Through explicit interaction, the user can directly feed the input or response to the AI system. 
It can be achieved through interfaces or controls, for example, by visually interacting with interface elements \cite{ramos2020interactive}.

\subsubsection{Augmentation and Control}
Augmentation, often a paradox with automation \cite{raisch2021artificial}, is the enhancement of abilities to perceive or achieve a desired task using AI systems. 
Automation, including the one using AI, strives to achieve autonomy to do the task on behalf of humans, taking control out of their hands.
Human-AI approaches advocate for building augmentative approaches that facilitate users rather than replacing them. 
Therefore, the control must remain with the user to exercise it and the user is always informed or aware of it.
Augmentation can work both ways: 1) AI enables users to enhance their understanding of the problem, and 2) human expertise improves the system's performance.
Augmentation is also defined as how the collaboration is carried out in human-AI design to reach the respective goals.

\subsubsection{Human-centered Design}
Human-computer interaction advocates for designing systems that match human cognition and mental models.
The human-centered design focuses on making systems more acceptable and benevolent.
The design of AI systems that cater to human factors such as their mental models, satisfaction, and expertise is considered human-centered, prioritizing user needs over algorithmic proficiency.
Various guidelines are proposed that can be employed in building human-centered design (i.e., by prioritizing user needs) \cite{amershi2019guidelines, googleaiguidebook}. 
For instance, design should include an explicit and informed understanding of the AI systems, their functionalities, and their limitations to end-users. 
Human-centered design can be attributed to an understanding of who the users are, the tasks they perform, and the objectives they want to achieve. 
For instance, many AI systems achieve this task by creating a better experience for the users to match their needs with expectations \cite{feng2023addressing}.

\section{Analysis of Dimensions}
\label{section:4analysis}
\begin{figure}[h]
  \centering
  \includegraphics[width=\linewidth]{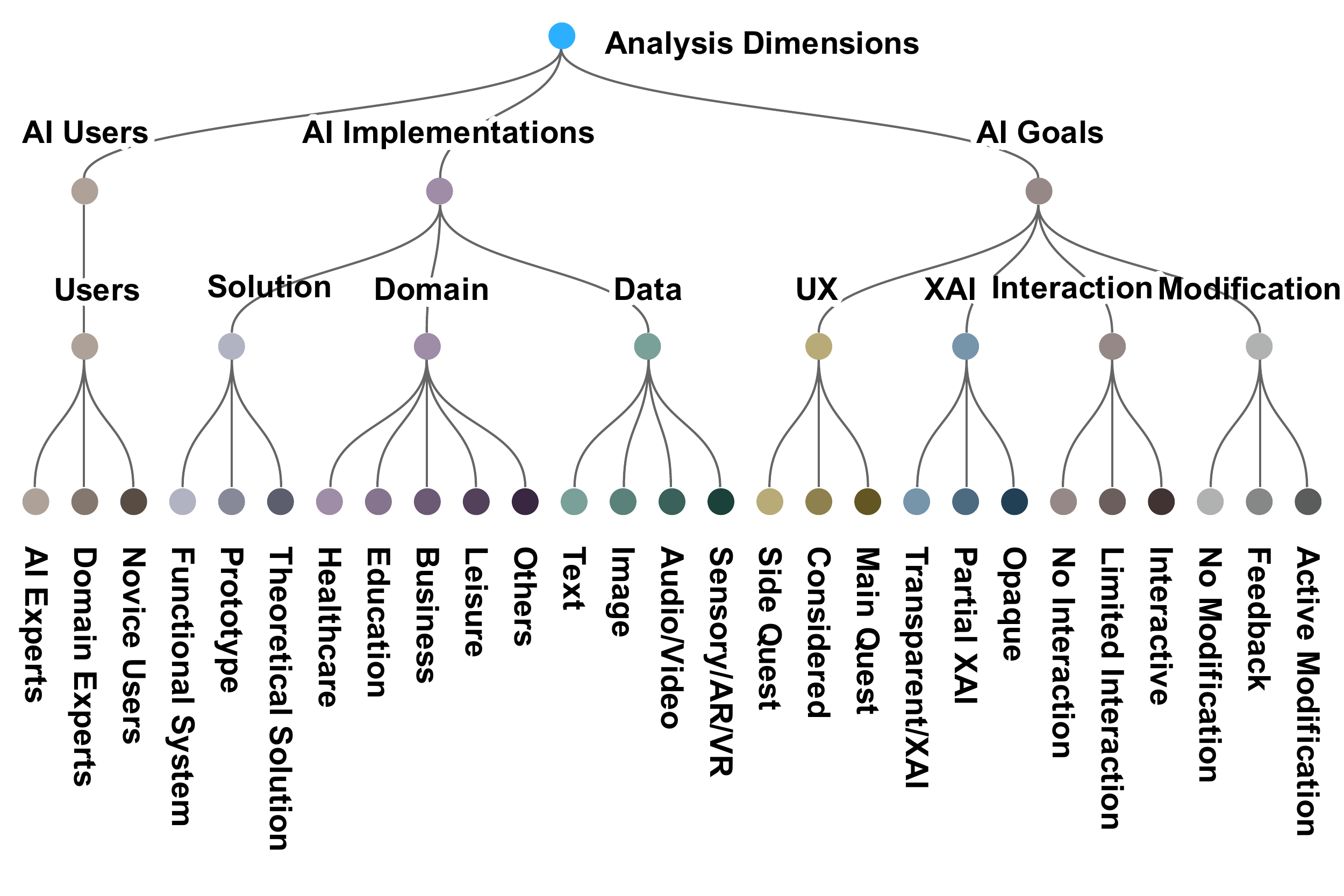}
  \caption{Categorization and overview of dimensions identified for the analysis. The major dimensions include users, implementations (implementation details, application domains, and data modalities), and goals (user experience, explainability, interaction, and active modification) of AI. Dimensions are further granulated to evaluate studies in detail.}
  \label{fig:summary_cats}
  % \Description{Tree diagram showing main areas of research agendas and topics (including sub-topics) to be examined.}
\end{figure}

Building on the methodology, we utilized selective coding to capture research attributes from the shortlisted studies.
We iteratively connected interrelated emerging patterns and analysis insights using a design thinking approach \cite{ac4d2023all}.
The details of captured attributes and analysis insights (dimensions) are listed in table \ref{tab:maintable}. 
Attributes were grouped in terms of users (AI, domain, novice users), data modality (text, image, A/V, sensors), implementation (solution, prototype, theory), application area (health, business, education, leisure), and study goals (user experience, transparency, interactivity, adaptation).
Attributes such as interactivity, human involvement, adaptations, and agency were reviewed cohesively.
Figure \ref{fig:summary_cats} shows an overall summary of analysis dimensions and their sub-categorizations.
Next, we evaluated the identified dimensions against stated objectives and contributions related to interactive AI.
We acknowledge the limitations of analysis as being based on the information available in publications’ content. 

\begin{sidewaystable}
  \caption{Description of unique publications along with dimensions and sub-categories for analysis of identified literature.}
  \label{tab:maintable}
  \scriptsize
  \setlength\tabcolsep{2pt}
  {\fontsize{7}{11}\selectfont
  \begin{tabular}{|l|p{4cm}|l|l|l|l|l|l|l|l|l|l|l|l|l|l|l|l|l|l|l|l|l|l|l|l|l|l|l|l|l|l|l|l|l|l|l|l|l|}
    \hline
    % & & & & \multicolumn{5}{|c|}{Users} & \multicolumn{3}{c|}{IMP*} & \multicolumn{6}{c|}{Domains} & \multicolumn{5}{c|}{Data Modality} & \multicolumn{3}{c|}{UX*} & \multicolumn{3}{c|}{TRNS*} & \multicolumn{3}{c|}{INTR*} & \multicolumn{3}{c|}{MDF*} & \multicolumn{2}{c|}{AG*} & \multicolumn{2}{c|}{MD*} \\ \cline{5-39}
    
    Sr. & Author(s) \& Study & Identified & Class & {\rotatebox[origin=c]{90}{AI Expert Users}} & {\rotatebox[origin=c]{90}{Domain Expert Users}} & {\rotatebox[origin=c]{90}{Novice Users}} & {\rotatebox[origin=c]{90}{Multi-User Group}} & {\rotatebox[origin=c]{90}{Unspecified}} & {\rotatebox[origin=c]{90}{Functional System}} & {\rotatebox[origin=c]{90}{Prototype (Low Functionality)}} & {\rotatebox[origin=c]{90}{Theoretical Solution or Proposal}} & {\rotatebox[origin=c]{90}{Healthcare Domain}} & {\rotatebox[origin=c]{90}{Education Domain}} & {\rotatebox[origin=c]{90}{Business Domain}} & {\rotatebox[origin=c]{90}{Leisure Domain}} & {\rotatebox[origin=c]{90}{Other Domains}} & {\rotatebox[origin=c]{90}{Unspecified}} & {\rotatebox[origin=c]{90}{Textual Data}} & {\rotatebox[origin=c]{90}{Image Data}} & {\rotatebox[origin=c]{90}{Audio/Video Data}} & {\rotatebox[origin=c]{90}{Sensory/AR/VR Data}} & {\rotatebox[origin=c]{90}{Data Unspecified}} & {\rotatebox[origin=c]{90}{Side Quest - UX}} & {\rotatebox[origin=c]{90}{Considered - UX}} & {\rotatebox[origin=c]{90}{Main Quest - UX}} & {\rotatebox[origin=c]{90}{Transparent/XAI}} & {\rotatebox[origin=c]{90}{Partial XAI}} & {\rotatebox[origin=c]{90}{Opaque AI}} & {\rotatebox[origin=c]{90}{No Interaction}} & {\rotatebox[origin=c]{90}{Limited Interaction}} & {\rotatebox[origin=c]{90}{Interactive}} & {\rotatebox[origin=c]{90}{No Modification}} & {\rotatebox[origin=c]{90}{Feedback Modification}} & {\rotatebox[origin=c]{90}{Active Modification}} & {\rotatebox[origin=c]{90}{User Augmentation}} & {\rotatebox[origin=c]{90}{System Augmentation}} & {\rotatebox[origin=c]{90}{No Modification}} & {\rotatebox[origin=c]{90}{Modification/Manipulation}} \\ \hline
    1 & Fails \& Olsen \cite{fails2003interactive} & Primary & IML & ~ & \cellcolor{ug_color!40}~ & ~ & ~ & ~ & \cellcolor{imp_color!30}~ & ~ & ~ & ~ & ~ & ~ & ~ & ~ & \cellcolor{dom_color} ~ & ~ & \cellcolor{data_color!40}~ & ~ & ~ & ~ & ~ & ~ & \cellcolor{ux_color}~ & ~ & ~ & \cellcolor{tr_color} ~ & ~ & ~ & \cellcolor{int_color}~ & ~ & \cellcolor{mod_color!60}~ & ~ & ~ & \cellcolor{ag_color}~ & ~ & \cellcolor{md_color}~ \\ \hline
    2 & Ruttkay et al. \cite{ruttkay2006human} & Snowballing & CollabAI & ~ & ~ & ~ & ~ & \cellcolor{ug_color} ~ & \cellcolor{imp_color!30} ~ & ~ & ~ & ~ & ~ & ~ & \cellcolor{dom_color!60} ~ & ~ & ~ & ~ & ~ & ~ & \cellcolor{data_color!80} ~ & ~ & ~ & ~ & \cellcolor{ux_color} ~ & \cellcolor{tr_color!30} ~ & ~ & ~ & ~ & ~ & \cellcolor{int_color} ~ & ~ & \cellcolor{mod_color!60} ~ & ~ & ~ & \cellcolor{ag_color} ~ & ~ & \cellcolor{md_color} ~ \\ \hline
    3 & Amershi et al. \cite{amershi2014power} & Primary & IML & ~ & ~ & ~ & ~ & \cellcolor{ug_color} ~ & ~ & ~ & \cellcolor{imp_color} ~ & ~ & ~ & ~ & ~ & ~ & \cellcolor{dom_color} ~ & ~ & ~ & ~ & ~ & \cellcolor{data_color} ~ & ~ & ~ & \cellcolor{ux_color} ~ & ~ & \cellcolor{tr_color!60} ~ & ~ & ~ & \cellcolor{int_color!60} ~ & ~ & ~ & \cellcolor{mod_color!60} ~ & ~ & \cellcolor{ag_color!50} ~ & ~ & ~ & \cellcolor{md_color} ~ \\ \hline
    4 & d’Inverno \& M. \cite{d2015heroic} & Primary & CollabAI & ~ & ~ & ~ & ~ & \cellcolor{ug_color} ~ & ~ & ~ & \cellcolor{imp_color} ~ & ~ & ~ & ~ & \cellcolor{dom_color!60} ~ & ~ & ~ & ~ & ~ & \cellcolor{data_color!60} ~ & ~ & ~ & ~ & ~ & \cellcolor{ux_color} ~ & ~ & ~ & \cellcolor{tr_color} ~ & ~ & ~ & \cellcolor{int_color} ~ & \cellcolor{mod_color!30} ~ & ~ & ~ & \cellcolor{ag_color!50} ~ & ~ & \cellcolor{md_color!50} ~ & ~ \\ \hline
    5 & Kulesza et al. \cite{kulesza2015principles} & Snowballing & CollabAI & ~ & ~ & ~ & \cellcolor{ug_color!80} ~ & ~ & \cellcolor{imp_color!30} ~ & ~ & ~ & ~ & ~ & ~ & \cellcolor{dom_color!60} ~ & ~ & ~ & \cellcolor{data_color!20} ~ & ~ & ~ & ~ & ~ & ~ & ~ & \cellcolor{ux_color} ~ & \cellcolor{tr_color!30} ~ & ~ & ~ & ~ & ~ & \cellcolor{int_color} ~ & ~ & \cellcolor{mod_color!60} ~ & ~ & ~ & \cellcolor{ag_color} ~ & ~ & \cellcolor{md_color} ~ \\ \hline        
    6 & Holzinger et al. \cite{holzinger2016towards} & Snowballing & CollabAI & \cellcolor{ug_color!20} ~ & ~ & ~ & ~ & ~ & \cellcolor{imp_color!30} ~ & ~ & ~ & ~ & ~ & \cellcolor{dom_color!45} ~ & ~ & ~ & ~ & ~ & \cellcolor{data_color!40} ~ & ~ & ~ & ~ & ~ & \cellcolor{ux_color!60} ~ & ~ & \cellcolor{tr_color!30} ~ & ~ & ~ & ~ & \cellcolor{int_color!60} ~ & ~ & ~ & \cellcolor{mod_color!60} ~ & ~ & ~ & \cellcolor{ag_color} ~ & ~ & \cellcolor{md_color} ~ \\ \hline
    7 & Kamar \cite{kamar2016directions} & Primary & HybridAI & ~ & ~ & ~ & ~ & \cellcolor{ug_color} ~ & ~ & ~ & \cellcolor{imp_color} ~ & ~ & ~ & ~ & ~ & ~ & \cellcolor{dom_color} ~ & ~ & ~ & ~ & ~ & \cellcolor{data_color} ~ & \cellcolor{ux_color!30} ~ & ~ & ~ & ~ & ~ & \cellcolor{tr_color} ~ & \cellcolor{int_color!30} ~ & ~ & ~ & \cellcolor{mod_color!30} ~ & ~ & ~ & ~ & \cellcolor{ag_color} ~ & \cellcolor{md_color!50} ~ & ~ \\ \hline
    8 & Koch \cite{koch2017design} & Primary & CollabAI & ~ & ~ & ~ & ~ & \cellcolor{ug_color} ~ & ~ & ~ & \cellcolor{imp_color} ~ & ~ & ~ & ~ & ~ & ~ & \cellcolor{dom_color} ~ & ~ & ~ & ~ & ~ & \cellcolor{data_color} ~ & ~ & \cellcolor{ux_color!60} ~ & ~ & ~ & ~ & \cellcolor{tr_color} ~ & ~ & \cellcolor{int_color!60} ~ & ~ & \cellcolor{mod_color!30} ~ & ~ & ~ & \cellcolor{ag_color!50} ~ & ~ & \cellcolor{md_color!50} ~ & ~ \\ \hline
    9 & Dellermann et al. \cite{dellermann2019hybrid} & Snowballing & HCAI & ~ & ~ & ~ & ~ & \cellcolor{ug_color} ~ & ~ & ~ & \cellcolor{imp_color} ~ & ~ & ~ & ~ & ~ & ~ & \cellcolor{dom_color} ~ & ~ & ~ & ~ & ~ & \cellcolor{data_color} ~ & \cellcolor{ux_color!30} ~ & ~ & ~ & ~ & ~ & \cellcolor{tr_color} ~ & \cellcolor{int_color!30} ~ & ~ & ~ & \cellcolor{mod_color!30} ~ & ~ & ~ & ~ & \cellcolor{ag_color} ~ & \cellcolor{md_color!50} ~ & ~ \\ \hline
    10 & Koch \& Oulasvirta \cite{koch2018group} & Primary & CollabAI & ~ & ~ & ~ & ~ & \cellcolor{ug_color} ~ & ~ & ~ & \cellcolor{imp_color} ~ & ~ & ~ & ~ & ~ & ~ & \cellcolor{dom_color} ~ & ~ & ~ & ~ & ~ & \cellcolor{data_color} ~ & ~ & \cellcolor{ux_color!60} ~ & ~ & ~ & \cellcolor{tr_color!60} ~ & ~ & ~ & \cellcolor{int_color!60} ~ & ~ & \cellcolor{mod_color!30} ~ & ~ & ~ & \cellcolor{ag_color!50} ~ & ~ & \cellcolor{md_color!50} ~ & ~ \\ \hline
    11 & Nicholls \cite{nicholls2018collaborative} & Primary & CollabAI & ~ & ~ & ~ & \cellcolor{ug_color!80} ~ & ~ & ~ & ~ & \cellcolor{imp_color} ~ & ~ & ~ & ~ & \cellcolor{dom_color!60} ~ & ~ & ~ & ~ & ~ & \cellcolor{data_color!60} ~ & ~ & ~ & \cellcolor{ux_color!30} ~ & ~ & ~ & ~ & ~ & \cellcolor{tr_color} ~ & ~ & ~ & \cellcolor{int_color} ~ & \cellcolor{mod_color!30} ~ & ~ & ~ & \cellcolor{ag_color!50} ~ & ~ & \cellcolor{md_color!50} ~ & ~ \\ \hline
    12 & Xie et al. \cite{xie2019vrgym} & Primary & InteractAI & ~ & ~ & ~ & \cellcolor{ug_color!80} ~ & ~ & \cellcolor{imp_color!30} ~ & ~ & ~ & ~ & \cellcolor{dom_color!30} ~ & ~ & ~ & ~ & ~ & ~ & ~ & ~ & \cellcolor{data_color!80} ~ & ~ & ~ & \cellcolor{ux_color!60} ~ & ~ & ~ & ~ & \cellcolor{tr_color} ~ & ~ & \cellcolor{int_color!60} ~ & ~ & \cellcolor{mod_color!30} ~ & ~ & ~ & ~ & \cellcolor{ag_color} ~ & \cellcolor{md_color!50} ~ & ~ \\ \hline
    13 & Patel \& Bhalodiya \cite{patel20193d} & Primary & InteractAI & \cellcolor{ug_color!20} ~ & ~ & ~ & ~ & ~ & \cellcolor{imp_color!30} ~ & ~ & ~ & ~ & ~ & ~ & ~ & ~ & \cellcolor{dom_color} ~ & ~ & ~ & ~ & \cellcolor{data_color!80} ~ & ~ & \cellcolor{ux_color!30} ~ & ~ & ~ & ~ & ~ & \cellcolor{tr_color} ~ & ~ & ~ & \cellcolor{int_color} ~ & \cellcolor{mod_color!30} ~ & ~ & ~ & ~ & \cellcolor{ag_color} ~ & \cellcolor{md_color!50} ~ & ~ \\ \hline
    14 & Liu et al. \cite{liu2019mappa} & Primary & InteractAI & ~ & ~ & ~ & \cellcolor{ug_color!80} ~ & ~ & \cellcolor{imp_color!30} ~ & ~ & ~ & ~ & \cellcolor{dom_color!30} ~ & ~ & ~ & ~ & ~ & ~ & ~ & \cellcolor{data_color!60} ~ & ~ & ~ & ~ & ~ & \cellcolor{ux_color} ~ & ~ & ~ & \cellcolor{tr_color} ~ & ~ & ~ & \cellcolor{int_color} ~ & ~ & ~ & \cellcolor{mod_color} ~ & \cellcolor{ag_color!50} ~ & ~ & ~ & \cellcolor{md_color} ~ \\ \hline
    15 & Heer \cite{heer2019agency} & Snowballing & HCAI & ~ & ~ & ~ & ~ & \cellcolor{ug_color} ~ & ~ & \cellcolor{imp_color!60} ~ & ~ & ~ & \cellcolor{dom_color!30} ~ & ~ & ~ & ~ & ~ & \cellcolor{data_color!20} ~ & ~ & ~ & ~ & ~ & ~ & \cellcolor{ux_color!60} ~ & ~ & ~ & ~ & \cellcolor{tr_color} ~ & ~ & \cellcolor{int_color!60} ~ & ~ & ~ & \cellcolor{mod_color!60} ~ & ~ & \cellcolor{ag_color!50} ~ & ~ & ~ & \cellcolor{md_color} ~ \\ \hline
    16 & Teso \& Kersting \cite{teso2019explanatory} & Primary & IML & \cellcolor{ug_color!20} ~ & ~ & ~ & ~ & ~ & ~ & ~ & \cellcolor{imp_color} ~ & ~ & ~ & ~ & ~ & ~ & \cellcolor{dom_color} ~ & \cellcolor{data_color!20} ~ & ~ & ~ & ~ & ~ & ~ & \cellcolor{ux_color!60} ~ & ~ & \cellcolor{tr_color!30} ~ & ~ & ~ & ~ & ~ & \cellcolor{int_color} ~ & ~ & \cellcolor{mod_color!60} ~ & ~ & ~ & \cellcolor{ag_color} ~ & ~ & \cellcolor{md_color} ~ \\ \hline
    17 & Cai et al. \cite{cai2019human} & Snowballing & IML & ~ & \cellcolor{ug_color!40} ~ & ~ & ~ & ~ & \cellcolor{imp_color!30} ~ & ~ & ~ & \cellcolor{dom_color!15} ~ & ~ & ~ & ~ & ~ & ~ & ~ & \cellcolor{data_color!40} ~ & ~ & ~ & ~ & ~ & ~ & \cellcolor{ux_color} ~ & ~ & ~ & \cellcolor{tr_color} ~ & ~ & ~ & \cellcolor{int_color} ~ & ~ & \cellcolor{mod_color!60} ~ & ~ & \cellcolor{ag_color!50} ~ & ~ & ~ & \cellcolor{md_color} ~ \\ \hline
    18 & Moradi et al. \cite{morteza2019collective} & Primary & HybridAI & ~ & ~ & ~ & ~ & \cellcolor{ug_color} ~ & ~ & ~ & \cellcolor{imp_color} ~ & ~ & ~ & ~ & ~ & ~ & \cellcolor{dom_color} ~ & ~ & ~ & ~ & ~ & \cellcolor{data_color} ~ & \cellcolor{ux_color!30} ~ & ~ & ~ & ~ & ~ & \cellcolor{tr_color} ~ & \cellcolor{int_color!30} ~ & ~ & ~ & \cellcolor{mod_color!30} ~ & ~ & ~ & ~ & \cellcolor{ag_color} ~ & \cellcolor{md_color!50} ~ & ~ \\ \hline
    19 & Sanfeliu Cortés et al. \cite{sanfeliu2020collaborative} & Primary & CollabAI & ~ & ~ & ~ & \cellcolor{ug_color!80} ~ & ~ & \cellcolor{imp_color!30} ~ & ~ & ~ & ~ & ~ & ~ & ~ & \cellcolor{dom_color!80} ~ & ~ & ~ & ~ & \cellcolor{data_color!60} ~ & ~ & ~ & \cellcolor{ux_color!30} ~ & ~ & ~ & ~ & ~ & \cellcolor{tr_color} ~ & ~ & ~ & \cellcolor{int_color} ~ & \cellcolor{mod_color!30} ~ & ~ & ~ & \cellcolor{ag_color!50} ~ & ~ & \cellcolor{md_color!50} ~ & ~ \\ \hline
    20 & Mruthyunjaya \& J. \cite{mruthyunjaya2020human} & Primary & CollabAI & ~ & ~ & ~ & \cellcolor{ug_color!80} ~ & ~ & ~ & \cellcolor{imp_color!60} ~ & ~ & ~ & ~ & ~ & ~ & \cellcolor{dom_color!80} ~ & ~ & ~ & ~ & ~ & \cellcolor{data_color!80} ~ & ~ & \cellcolor{ux_color!30} ~ & ~ & ~ & ~ & ~ & \cellcolor{tr_color} ~ & ~ & \cellcolor{int_color!60} ~ & ~ & ~ & \cellcolor{mod_color!60} ~ & ~ & ~ & \cellcolor{ag_color} ~ & ~ & \cellcolor{md_color} ~ \\ \hline
    21 & McCormack et al. \cite{mccormack2020design} & Snowballing & CollabAI & \cellcolor{ug_color!20} ~ & ~ & ~ & ~ & ~ & ~ & ~ & \cellcolor{imp_color} ~ & ~ & ~ & ~ & \cellcolor{dom_color!60} ~ & ~ & ~ & ~ & ~ & ~ & ~ & \cellcolor{data_color} ~ & ~ & ~ & \cellcolor{ux_color} ~ & ~ & ~ & \cellcolor{tr_color} ~ & ~ & ~ & \cellcolor{int_color} ~ & ~ & \cellcolor{mod_color!60} ~ & ~ & \cellcolor{ag_color!50} ~ & ~ & ~ & \cellcolor{md_color} ~ \\ \hline
    22 & Kambhampati \cite{kambhampati2020challenges} & Snowballing & CollabAI & ~ & ~ & ~ & ~ & \cellcolor{ug_color} ~ & ~ & ~ & \cellcolor{imp_color} ~ & ~ & ~ & ~ & ~ & \cellcolor{dom_color!80} ~ & ~ & ~ & ~ & ~ & ~ & \cellcolor{data_color} ~ & ~ & \cellcolor{ux_color!60} ~ & ~ & ~ & \cellcolor{tr_color!60} ~ & ~ & ~ & \cellcolor{int_color!60} ~ & ~ & \cellcolor{mod_color!30} ~ & ~ & ~ & \cellcolor{ag_color!50} ~ & ~ & \cellcolor{md_color!50} ~ & ~ \\ \hline
    23 & Yang et al. \cite{yang2020hybrid} & Primary & InteractAI & ~ & ~ & \cellcolor{ug_color!60} ~ & ~ & ~ & \cellcolor{imp_color!30} ~ & ~ & ~ & ~ & \cellcolor{dom_color!30} ~ & ~ & ~ & ~ & ~ & ~ & ~ & \cellcolor{data_color!60} ~ & ~ & ~ & \cellcolor{ux_color!30} ~ & ~ & ~ & ~ & ~ & \cellcolor{tr_color} ~ & ~ & ~ & \cellcolor{int_color} ~ & \cellcolor{mod_color!30} ~ & ~ & ~ & \cellcolor{ag_color!50} ~ & ~ & \cellcolor{md_color!50} ~ & ~ \\ \hline
    24 & Dotov \& Froese \cite{dotov2020dynamic} & Primary & InteractAI & \cellcolor{ug_color!20} ~ & ~ & ~ & ~ & ~ & ~ & \cellcolor{imp_color!60} ~ & ~ & ~ & ~ & ~ & \cellcolor{dom_color!60} ~ & ~ & ~ & ~ & ~ & ~ & ~ & \cellcolor{data_color} ~ & \cellcolor{ux_color!30} ~ & ~ & ~ & ~ & ~ & \cellcolor{tr_color} ~ & \cellcolor{int_color!30} ~ & ~ & ~ & \cellcolor{mod_color!30} ~ & ~ & ~ & \cellcolor{ag_color!50} ~ & ~ & \cellcolor{md_color!50} ~ & ~ \\ \hline
    25 & Shneiderman \cite{shneiderman2020human} & Primary & HCAI & ~ & ~ & ~ & ~ & \cellcolor{ug_color} ~ & ~ & ~ & \cellcolor{imp_color} ~ & ~ & ~ & ~ & ~ & ~ & \cellcolor{dom_color} ~ & ~ & ~ & ~ & ~ & \cellcolor{data_color} ~ & ~ & ~ & \cellcolor{ux_color} ~ & ~ & ~ & \cellcolor{tr_color} ~ & ~ & ~ & \cellcolor{int_color} ~ & ~ & \cellcolor{mod_color!60} ~ & ~ & \cellcolor{ag_color!50} ~ & ~ & ~ & \cellcolor{md_color} ~ \\ \hline
    26 & Shneiderman \cite{shneiderman2020human_rel} & Primary & HCAI & ~ & ~ & ~ & ~ & \cellcolor{ug_color} ~ & ~ & ~ & \cellcolor{imp_color} ~ & ~ & ~ & ~ & ~ & ~ & \cellcolor{dom_color} ~ & ~ & ~ & ~ & ~ & \cellcolor{data_color} ~ & ~ & ~ & \cellcolor{ux_color} ~ & ~ & ~ & \cellcolor{tr_color} ~ & ~ & ~ & \cellcolor{int_color} ~ & ~ & \cellcolor{mod_color!60} ~ & ~ & \cellcolor{ag_color!50} ~ & ~ & ~ & \cellcolor{md_color} ~ \\ \hline
    27 & Auernhammer \cite{auernhammer2020human} & Primary & HCAI & ~ & ~ & ~ & ~ & \cellcolor{ug_color} ~ & ~ & ~ & \cellcolor{imp_color} ~ & ~ & ~ & \cellcolor{dom_color!45} ~ & ~ & ~ & ~ & ~ & ~ & ~ & ~ & \cellcolor{data_color} ~ & \cellcolor{ux_color!30} ~ & ~ & ~ & ~ & ~ & \cellcolor{tr_color} ~ & ~ & \cellcolor{int_color!60} ~ & ~ & \cellcolor{mod_color!30} ~ & ~ & ~ & \cellcolor{ag_color!50} ~ & ~ & \cellcolor{md_color!50} ~ & ~ \\ \hline
    28 & Honeycutt et al. \cite{honeycutt2020soliciting} & Primary & IML & ~ & ~ & ~ & ~ & \cellcolor{ug_color} ~ & \cellcolor{imp_color!30} ~ & ~ & ~ & ~ & \cellcolor{dom_color!30} ~ & ~ & ~ & ~ & ~ & ~ & \cellcolor{data_color!40} ~ & ~ & ~ & ~ & \cellcolor{ux_color!30} ~ & ~ & ~ & ~ & ~ & \cellcolor{tr_color} ~ & \cellcolor{int_color!30} ~ & ~ & ~ & \cellcolor{mod_color!30} ~ & ~ & ~ & \cellcolor{ag_color!50} ~ & ~ & \cellcolor{md_color!50} ~ & ~ \\ \hline
    
  \end{tabular} }
\end{sidewaystable}

\begin{sidewaystable}
  \ContinuedFloat
  \caption{(continued) Description of unique publications along with dimensions and sub-categories for analysis of identified literature.}
  % \label{tab:searchterms}
  \scriptsize
  \setlength\tabcolsep{2pt}
   {\fontsize{7}{11}\selectfont
  \begin{tabular}{|l|l|l|l|l|l|l|l|l|l|l|l|l|l|l|l|l|l|l|l|l|l|l|l|l|l|l|l|l|l|l|l|l|l|l|l|l|l|l|}
    \hline
    % & & & & \multicolumn{5}{|c|}{Users} & \multicolumn{3}{c|}{IMP*} & \multicolumn{6}{c|}{Domains} & \multicolumn{5}{c|}{Data Modality} & \multicolumn{3}{c|}{UX*} & \multicolumn{3}{c|}{TRNS*} & \multicolumn{3}{c|}{INTR*} & \multicolumn{3}{c|}{MDF*} & \multicolumn{2}{c|}{AG*} & \multicolumn{2}{c|}{MD*} \\ \cline{5-39}
    
    Sr. & Author(s) \& Study & Identified & Class & {\rotatebox[origin=c]{90}{AI Expert Users}} & {\rotatebox[origin=c]{90}{Domain Expert Users}} & {\rotatebox[origin=c]{90}{Novice Users}} & {\rotatebox[origin=c]{90}{Multi-User Group}} & {\rotatebox[origin=c]{90}{Unspecified}} & {\rotatebox[origin=c]{90}{Functional System}} & {\rotatebox[origin=c]{90}{Prototype (Low Functionality)}} & {\rotatebox[origin=c]{90}{Theoretical Solution or Proposal}} & {\rotatebox[origin=c]{90}{Healthcare Domain}} & {\rotatebox[origin=c]{90}{Education Domain}} & {\rotatebox[origin=c]{90}{Business Domain}} & {\rotatebox[origin=c]{90}{Leisure Domain}} & {\rotatebox[origin=c]{90}{Other Domains}} & {\rotatebox[origin=c]{90}{Unspecified}} & {\rotatebox[origin=c]{90}{Textual Data}} & {\rotatebox[origin=c]{90}{Image Data}} & {\rotatebox[origin=c]{90}{Audio/Video Data}} & {\rotatebox[origin=c]{90}{Sensory/AR/VR Data}} & {\rotatebox[origin=c]{90}{Data Unspecified}} & {\rotatebox[origin=c]{90}{Side Quest - UX}} & {\rotatebox[origin=c]{90}{Considered - UX}} & {\rotatebox[origin=c]{90}{Main Quest - UX}} & {\rotatebox[origin=c]{90}{Transparent/XAI}} & {\rotatebox[origin=c]{90}{Partial XAI}} & {\rotatebox[origin=c]{90}{Opaque AI}} & {\rotatebox[origin=c]{90}{No Interaction}} & {\rotatebox[origin=c]{90}{Limited Interaction}} & {\rotatebox[origin=c]{90}{Interactive}} & {\rotatebox[origin=c]{90}{No Modification}} & {\rotatebox[origin=c]{90}{Feedback Modification}} & {\rotatebox[origin=c]{90}{Active Modification}} & {\rotatebox[origin=c]{90}{User Augmentation}} & {\rotatebox[origin=c]{90}{System Augmentation}} & {\rotatebox[origin=c]{90}{No Modification}} & {\rotatebox[origin=c]{90}{Modification/Manipulation}} \\ \hline
    29 & Carney et al. \cite{carney2020teachable} & Primary & IML & ~ & \cellcolor{ug_color!40} ~ & ~ & ~ & ~ & \cellcolor{imp_color!30} ~ & ~ & ~ & ~ & ~ & ~ & ~ & ~ & \cellcolor{dom_color} ~ & ~ & \cellcolor{data_color!40} ~ & ~ & ~ & ~ & ~ & ~ & \cellcolor{ux_color} ~ & ~ & ~ & \cellcolor{tr_color} ~ & ~ & ~ & \cellcolor{int_color} ~ & ~ & \cellcolor{mod_color!60} ~ & ~ & \cellcolor{ag_color!50} ~ & ~ & ~ & \cellcolor{md_color} ~ \\ \hline
    30 & Wexler et al. \cite{wexler2019if} & Primary & IML & \cellcolor{ug_color!20} ~ & ~ & ~ & ~ & ~ & \cellcolor{imp_color!30} ~ & ~ & ~ & ~ & ~ & ~ & ~ & ~ & \cellcolor{dom_color} ~ & \cellcolor{data_color!20} ~ & ~ & ~ & ~ & ~ & ~ & \cellcolor{ux_color!60} ~ & ~ & ~ & ~ & \cellcolor{tr_color} ~ & ~ & ~ & \cellcolor{int_color} ~ & ~ & \cellcolor{mod_color!60} ~ & ~ & \cellcolor{ag_color!50} ~ & ~ & ~ & \cellcolor{md_color} ~ \\ \hline
    31 & Ramos et al. \cite{ramos2020interactive} & Primary & IML & ~ & ~ & ~ & \cellcolor{ug_color!80} ~ & ~ & ~ & \cellcolor{imp_color!60} ~ & ~ & ~ & ~ & \cellcolor{dom_color!45} ~ & ~ & ~ & ~ & \cellcolor{data_color!20} ~ & ~ & ~ & ~ & ~ & ~ & \cellcolor{ux_color!60} ~ & ~ & \cellcolor{tr_color!30} ~ & ~ & ~ & ~ & ~ & \cellcolor{int_color} ~ & ~ & \cellcolor{mod_color!60} ~ & ~ & \cellcolor{ag_color!50} ~ & ~ & ~ & \cellcolor{md_color} ~ \\ \hline
    32 & Peeters et al. \cite{peeters2021hybrid} & Primary & HybridAI & ~ & ~ & ~ & ~ & \cellcolor{ug_color} ~ & ~ & ~ & \cellcolor{imp_color} ~ & ~ & ~ & ~ & ~ & ~ & \cellcolor{dom_color} ~ & ~ & ~ & ~ & ~ & \cellcolor{data_color} ~ & \cellcolor{ux_color!30} ~ & ~ & ~ & ~ & ~ & \cellcolor{tr_color} ~ & ~ & \cellcolor{int_color!60} ~ & ~ & \cellcolor{mod_color!30} ~ & ~ & ~ & ~ & \cellcolor{ag_color} ~ & \cellcolor{md_color!50} ~ & ~ \\ \hline
    33 & Agostinelli et al. \cite{agostinelli2021designing} & Primary & CollabAI & ~ & ~ & \cellcolor{ug_color!60} ~ & ~ & ~ & ~ & \cellcolor{imp_color!60} ~ & ~ & ~ & \cellcolor{dom_color!30} ~ & ~ & ~ & ~ & ~ & \cellcolor{data_color!20} ~ & ~ & ~ & ~ & ~ & ~ & ~ & \cellcolor{ux_color} ~ & ~ & ~ & \cellcolor{tr_color} ~ & ~ & ~ & \cellcolor{int_color} ~ & ~ & \cellcolor{mod_color!60} ~ & ~ & \cellcolor{ag_color!50} ~ & ~ & ~ & \cellcolor{md_color} ~ \\ \hline
    34 & Fügener et al. \cite{fugener2021will} & Snowballing & CollabAI & ~ & ~ & ~ & ~ & \cellcolor{ug_color} ~ & ~ & \cellcolor{imp_color!60} ~ & ~ & ~ & ~ & ~ & ~ & ~ & \cellcolor{dom_color} ~ & ~ & \cellcolor{data_color!40} ~ & ~ & ~ & ~ & ~ & \cellcolor{ux_color!60} ~ & ~ & ~ & ~ & \cellcolor{tr_color} ~ & ~ & \cellcolor{int_color!60} ~ & ~ & \cellcolor{mod_color!30} ~ & ~ & ~ & \cellcolor{ag_color!50} ~ & ~ & \cellcolor{md_color!50} ~ & ~ \\ \hline
    35 & Puntoni et al. \cite{puntoni2021consumers} & Snowballing & CollabAI & ~ & ~ & ~ & \cellcolor{ug_color!80} ~ & ~ & ~ & ~ & \cellcolor{imp_color} ~ & ~ & ~ & \cellcolor{dom_color!45} ~ & ~ & ~ & ~ & ~ & ~ & ~ & ~ & \cellcolor{data_color} ~ & ~ & ~ & \cellcolor{ux_color} ~ & ~ & ~ & \cellcolor{tr_color} ~ & ~ & ~ & \cellcolor{int_color} ~ & \cellcolor{mod_color!30} ~ & ~ & ~ & \cellcolor{ag_color!50} ~ & ~ & \cellcolor{md_color!50} ~ & ~ \\ \hline
    36 & Xu et al. \cite{xu2023transitioning} & Primary & HCAI & ~ & ~ & ~ & ~ & \cellcolor{ug_color} ~ & ~ & ~ & \cellcolor{imp_color} ~ & ~ & ~ & ~ & ~ & ~ & \cellcolor{dom_color} ~ & ~ & ~ & ~ & ~ & \cellcolor{data_color} ~ & \cellcolor{ux_color!30} ~ & ~ & ~ & ~ & ~ & \cellcolor{tr_color} ~ & \cellcolor{int_color!30} ~ & ~ & ~ & \cellcolor{mod_color!30} ~ & ~ & ~ & \cellcolor{ag_color!50} ~ & ~ & \cellcolor{md_color!50} ~ & ~ \\ \hline
    37 & Urban Davis et al. \cite{urban2021designing} & Snowballing & HCAI & ~ & \cellcolor{ug_color!40} ~ & ~ & ~ & ~ & \cellcolor{imp_color!30} ~ & ~ & ~ & ~ & ~ & ~ & \cellcolor{dom_color!60} ~ & ~ & ~ & ~ & ~ & ~ & \cellcolor{data_color!80} ~ & ~ & ~ & ~ & \cellcolor{ux_color} ~ & ~ & ~ & \cellcolor{tr_color} ~ & ~ & ~ & \cellcolor{int_color} ~ & ~ & \cellcolor{mod_color!60} ~ & ~ & \cellcolor{ag_color!50} ~ & ~ & ~ & \cellcolor{md_color} ~ \\ \hline
    38 & Chignell et al. \cite{chignell2021human} & Primary & IML & ~ & \cellcolor{ug_color!40} ~ & ~ & ~ & ~ & \cellcolor{imp_color!30} ~ & ~ & ~ & ~ & ~ & ~ & ~ & \cellcolor{dom_color!80} ~ & ~ & \cellcolor{data_color!20} ~ & ~ & ~ & ~ & ~ & ~ & ~ & \cellcolor{ux_color} ~ & ~ & ~ & \cellcolor{tr_color} ~ & ~ & ~ & \cellcolor{int_color} ~ & ~ & \cellcolor{mod_color!60} ~ & ~ & ~ & \cellcolor{ag_color} ~ & ~ & \cellcolor{md_color} ~ \\ \hline
    39 & Mishra \& R. \cite{mishra2021designing} & Snowballing & IML & ~ & ~ & \cellcolor{ug_color!60} ~ & ~ & ~ & ~ & \cellcolor{imp_color!60} ~ & ~ & ~ & \cellcolor{dom_color!30} ~ & ~ & ~ & ~ & ~ & \cellcolor{data_color!20} ~ & ~ & ~ & ~ & ~ & ~ & ~ & \cellcolor{ux_color} ~ & ~ & \cellcolor{tr_color!60} ~ & ~ & ~ & \cellcolor{int_color!60} ~ & ~ & \cellcolor{mod_color!30} ~ & ~ & ~ & \cellcolor{ag_color!50} ~ & ~ & \cellcolor{md_color!50} ~ & ~ \\ \hline
    40 & Huang \& Rust \cite{huang2022framework} & Primary & CollabAI & ~ & ~ & ~ & ~ & \cellcolor{ug_color} ~ & ~ & ~ & \cellcolor{imp_color} ~ & ~ & ~ & \cellcolor{dom_color!45} ~ & ~ & ~ & ~ & \cellcolor{data_color!20} ~ & ~ & ~ & ~ & ~ & \cellcolor{ux_color!30} ~ & ~ & ~ & ~ & ~ & \cellcolor{tr_color} ~ & \cellcolor{int_color!30} ~ & ~ & ~ & ~ & ~ & \cellcolor{mod_color} ~ & ~ & \cellcolor{ag_color} ~ & ~ & \cellcolor{md_color} ~ \\ \hline
    41 & Ottoboni et al. \cite{ottoboni2022multifunctional} & Primary & InteractAI & ~ & \cellcolor{ug_color!40} ~ & ~ & ~ & ~ & ~ & ~ & \cellcolor{imp_color} ~ & \cellcolor{dom_color!15} ~ & ~ & ~ & ~ & ~ & ~ & ~ & ~ & ~ & ~ & \cellcolor{data_color} ~ & ~ & ~ & \cellcolor{ux_color} ~ & ~ & ~ & \cellcolor{tr_color} ~ & ~ & \cellcolor{int_color!60} ~ & ~ & \cellcolor{mod_color!30} ~ & ~ & ~ & \cellcolor{ag_color!50} ~ & ~ & \cellcolor{md_color!50} ~ & ~ \\ \hline
    42 & Doncieux et al. \cite{doncieux2022human} & Primary & HCAI & ~ & ~ & ~ & ~ & \cellcolor{ug_color} ~ & ~ & ~ & \cellcolor{imp_color} ~ & ~ & ~ & ~ & ~ & \cellcolor{dom_color!80} ~ & ~ & ~ & ~ & ~ & ~ & \cellcolor{data_color} ~ & ~ & \cellcolor{ux_color!60} ~ & ~ & ~ & ~ & \cellcolor{tr_color} ~ & ~ & ~ & \cellcolor{int_color} ~ & ~ & ~ & \cellcolor{mod_color} ~ & \cellcolor{ag_color!50} ~ & ~ & ~ & \cellcolor{md_color} ~ \\ \hline
    43 & Andersen et al. \cite{andersen2022integrating} & Primary & HCAI & ~ & ~ & \cellcolor{ug_color!60} ~ & ~ & ~ & \cellcolor{imp_color!30} ~ & ~ & ~ & ~ & \cellcolor{dom_color!30} ~ & ~ & ~ & ~ & ~ & \cellcolor{data_color!20} ~ & ~ & ~ & ~ & ~ & ~ & ~ & \cellcolor{ux_color} ~ & ~ & ~ & \cellcolor{tr_color} ~ & ~ & ~ & \cellcolor{int_color} ~ & \cellcolor{mod_color!30} ~ & ~ & ~ & \cellcolor{ag_color!50} ~ & ~ & ~ & \cellcolor{md_color} ~ \\ \hline
    44 & Nakao et al. \cite{nakao2022toward} & Primary & IML & ~ & ~ & ~ & \cellcolor{ug_color!80} ~ & ~ & ~ & \cellcolor{imp_color!60} ~ & ~ & ~ & ~ & ~ & ~ & \cellcolor{dom_color!80} ~ & ~ & \cellcolor{data_color!20} ~ & ~ & ~ & ~ & ~ & ~ & ~ & \cellcolor{ux_color} ~ & \cellcolor{tr_color!30} ~ & ~ & ~ & ~ & ~ & \cellcolor{int_color} ~ & ~ & \cellcolor{mod_color!60} ~ & ~ & ~ & \cellcolor{ag_color} ~ & ~ & \cellcolor{md_color} ~ \\ \hline
    45 & Lee et al. \cite{lee2022coauthor} & Snowballing & IML & ~ & ~ & \cellcolor{ug_color!60} ~ & ~ & ~ & \cellcolor{imp_color!30} ~ & ~ & ~ & ~ & \cellcolor{dom_color!30} ~ & ~ & ~ & ~ & ~ & \cellcolor{data_color!20} ~ & ~ & ~ & ~ & ~ & ~ & ~ & \cellcolor{ux_color} ~ & ~ & ~ & \cellcolor{tr_color} ~ & ~ & ~ & \cellcolor{int_color} ~ & \cellcolor{mod_color!30} ~ & ~ & ~ & \cellcolor{ag_color!50} ~ & ~ & \cellcolor{md_color!50} ~ & ~ \\ \hline
    46 & Wellsandt et al. \cite{wellsandt2022hybrid} & Primary & HybridAI & ~ & ~ & ~ & ~ & \cellcolor{ug_color} ~ & ~ & ~ & \cellcolor{imp_color} ~ & ~ & ~ & ~ & ~ & ~ & \cellcolor{dom_color} ~ & ~ & ~ & ~ & ~ & \cellcolor{data_color} ~ & \cellcolor{ux_color!30} ~ & ~ & ~ & ~ & ~ & \cellcolor{tr_color} ~ & \cellcolor{int_color!30} ~ & ~ & ~ & \cellcolor{mod_color!30} ~ & ~ & ~ & ~ & \cellcolor{ag_color} ~ & \cellcolor{md_color!50} ~ & ~ \\ \hline
    47 & Cabitza et al. \cite{cabitza2023rams} & Primary & ContestAI & ~ & \cellcolor{ug_color!40} ~ & ~ & ~ & ~ & \cellcolor{imp_color!30} ~ & ~ & ~ & \cellcolor{dom_color!15} ~ & ~ & ~ & ~ & ~ & ~ & \cellcolor{data_color!20} ~ & ~ & ~ & ~ & ~ & ~ & \cellcolor{ux_color!60} ~ & ~ & ~ & \cellcolor{tr_color!60} ~ & ~ & \cellcolor{int_color!30} ~ & ~ & ~ & \cellcolor{mod_color!30} ~ & ~ & ~ & \cellcolor{ag_color!50} ~ & ~ & \cellcolor{md_color!50} ~ & ~ \\ \hline
    48 & Maiden et al. \cite{maiden2023designing} & Primary & HCAI & ~ & \cellcolor{ug_color!40} ~ & ~ & ~ & ~ & \cellcolor{imp_color!30} ~ & ~ & ~ & ~ & ~ & ~ & \cellcolor{dom_color!60} ~ & ~ & ~ & \cellcolor{data_color!20} ~ & ~ & ~ & ~ & ~ & ~ & \cellcolor{ux_color!60} ~ & ~ & ~ & ~ & \cellcolor{tr_color} ~ & ~ & \cellcolor{int_color!60} ~ & ~ & \cellcolor{mod_color!30} ~ & ~ & ~ & \cellcolor{ag_color!50} ~ & ~ & \cellcolor{md_color!50} ~ & ~ \\ \hline
    49 & Feng and M. \cite{feng2023addressing} & Primary & IML & ~ & \cellcolor{ug_color!40} ~ & ~ & ~ & ~ & \cellcolor{imp_color!30} ~ & ~ & ~ & ~ & ~ & ~ & \cellcolor{dom_color!60} ~ & ~ & ~ & \cellcolor{data_color!20} ~ & ~ & ~ & ~ & ~ & ~ & \cellcolor{ux_color!60} ~ & ~ & ~ & ~ & \cellcolor{tr_color} ~ & ~ & \cellcolor{int_color!60} ~ & ~ & ~ & \cellcolor{mod_color!60} ~ & ~ & \cellcolor{ag_color!50} ~ & ~ & ~ & \cellcolor{md_color} ~ \\ \hline
    50 & Raees et al. \cite{raees2023four} & Primary & IML & ~ & \cellcolor{ug_color!40} ~ & ~ & ~ & ~ & ~ & \cellcolor{imp_color!60} ~ & ~ & ~ & ~ & \cellcolor{dom_color!45} ~ & ~ & ~ & ~ & \cellcolor{data_color!20} ~ & ~ & ~ & ~ & ~ & ~ & \cellcolor{ux_color!60} ~ & ~ & ~ & ~ & \cellcolor{tr_color} ~ & ~ & ~ & \cellcolor{int_color} ~ & ~ & \cellcolor{mod_color!60} ~ & ~ & \cellcolor{ag_color!50} ~ & ~ & ~ & \cellcolor{md_color} ~ \\ \hline
    51 & Kuang et al. \cite{kuang2023collaboration} & Primary & IML & ~ & \cellcolor{ug_color!40} ~ & ~ & ~ & ~ & ~ & \cellcolor{imp_color!60} ~ & ~ & ~ & ~ & \cellcolor{dom_color!45} ~ & ~ & ~ & ~ & \cellcolor{data_color!20} ~ & ~ & ~ & ~ & ~ & ~ & ~ & \cellcolor{ux_color} ~ & ~ & ~ & \cellcolor{tr_color} ~ & ~ & ~ & \cellcolor{int_color} ~ & \cellcolor{mod_color!30} ~ & ~ & ~ & \cellcolor{ag_color!50} ~ & ~ & \cellcolor{md_color!50} ~ & ~ \\ \hline
    52 & Chen et al. \cite{chen2023codeml} & Primary & IML & ~ & \cellcolor{ug_color!40} ~ & ~ & ~ & ~ & \cellcolor{imp_color!30} ~ & ~ & ~ & ~ & \cellcolor{dom_color!30} ~ & ~ & ~ & ~ & ~ & \cellcolor{data_color!20} ~ & ~ & ~ & ~ & ~ & ~ & \cellcolor{ux_color!60} ~ & ~ & ~ & ~ & \cellcolor{tr_color} ~ & ~ & \cellcolor{int_color!60} ~ & ~ & ~ & \cellcolor{mod_color!60} ~ & ~ & \cellcolor{ag_color!50} ~ & ~ & ~ & \cellcolor{md_color} ~ \\ \hline
    53 & Chiang et al. \cite{chiang2023two} & Snowballing & HybridAI & ~ & \cellcolor{ug_color!40} ~ & ~ & ~ & ~ & \cellcolor{imp_color!30} ~ & ~ & ~ & ~ & ~ & ~ & ~ & ~ & \cellcolor{dom_color} ~ & \cellcolor{data_color!20} ~ & ~ & ~ & ~ & ~ & ~ & ~ & \cellcolor{ux_color} ~ & ~ & \cellcolor{tr_color!60} ~ & ~ & ~ & ~ & \cellcolor{int_color} ~ & \cellcolor{mod_color!30} ~ & ~ & ~ & ~ & \cellcolor{ag_color} ~ & \cellcolor{md_color!50} ~ & ~ \\ \hline
    54 & Inkpen et al. \cite{inkpen2023advancing} & Snowballing & HybridAI & ~ & \cellcolor{ug_color!40} ~ & ~ & ~ & ~ & \cellcolor{imp_color!30} ~ & ~ & ~ & \cellcolor{dom_color!15} ~ & ~ & ~ & ~ & ~ & ~ & ~ & ~ & \cellcolor{data_color!60} ~ & ~ & ~ & ~ & \cellcolor{ux_color!60} ~ & ~ & ~ & ~ & \cellcolor{tr_color} ~ & ~ & ~ & \cellcolor{int_color} ~ & ~ & \cellcolor{mod_color!60} ~ & ~ & \cellcolor{ag_color!50} ~ & ~ & ~ & \cellcolor{md_color} ~ \\ \hline
    ~ & ~ & ~ & Count & 6 & 14 & 5 & 9 & 20 & 23 & 10 & 21 & 4 & 10 & 7 & 9 & 6 & 18 & 19 & 6 & 6 & 5 & 18 & 15 & 17 & 22 & 6 & 6 & 42 & 9 & 16 & 29 & 28 & 23 & 3 & 37 & 17 & 27 & 27 \\ \hline
  \end{tabular} }
\end{sidewaystable}

\subsection{AI Users}
Interaction is a driving factor in transforming traditional AI into user-centered AI making the user dimension an essential parameter to gauge the applicability of systems for a target audience.
However, isolated AI applications that restrain user participation affect interactivity and user perception about being part of the process.
In interactivity studies, user testing and participation is the central theme, however, it varies depending on the recruitment methods, type of task, and the expertise of users.
AI systems are targeted at diverse user groups ranging from AI experts to novices.
Various factors, such as the type of application, participants' availability, resources to recruit, and the complexity of AI systems, affect the selection of user groups. 
Hence, a clear profiling of target users is essential for human-machine interactions before developing AI systems.
Generally, users are grouped into three categories, namely, practitioners, domain experts, and novices, based on their skills and knowledge. 
Practitioners are people who are experts in building and using AI/ML systems, domain experts are the people who are experts in the problem domain, and novices are users without much knowledge about the AI concepts or problem domain.

\begin{figure}[h]
  \centering
  \includegraphics[height=5cm]{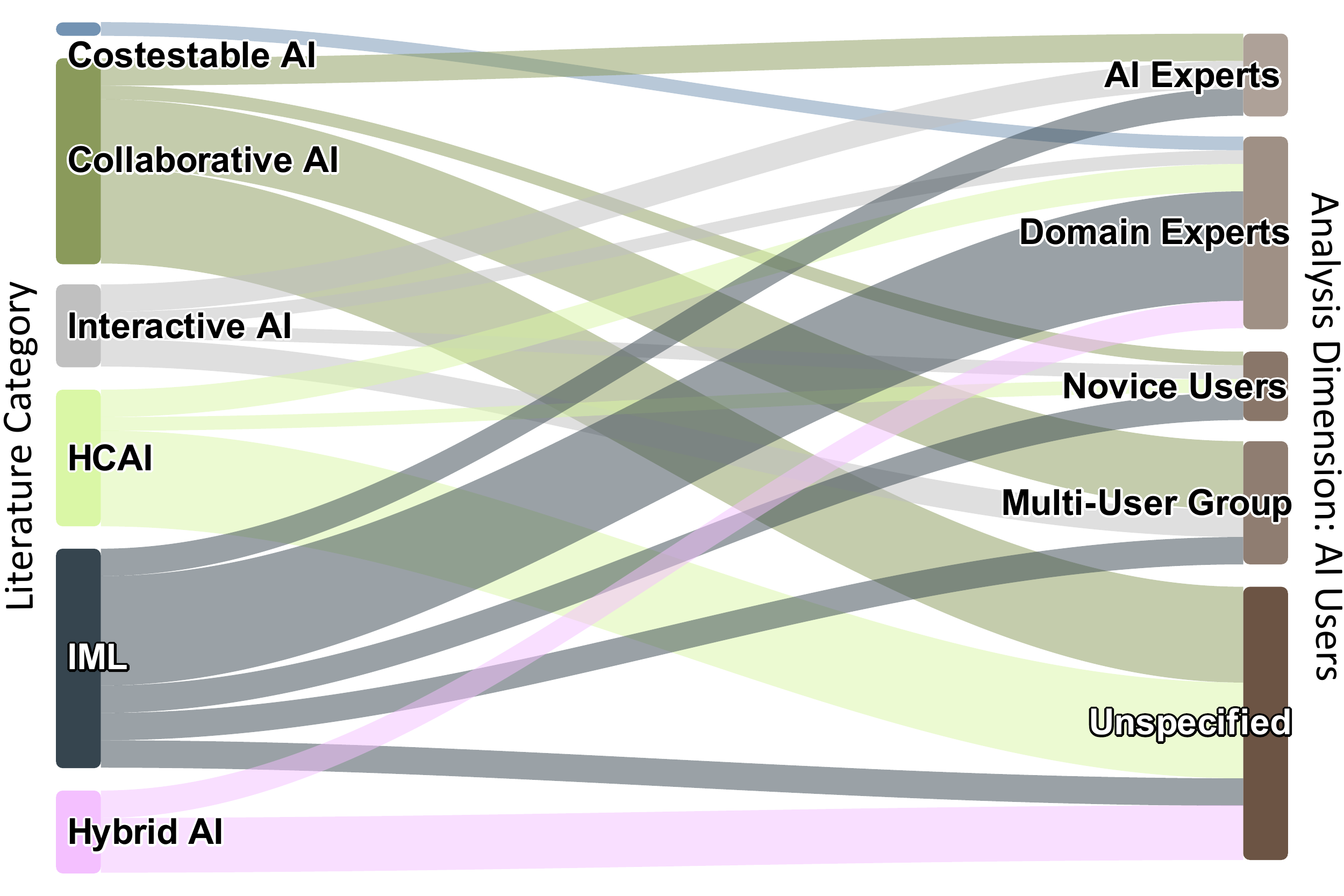}
  \caption{Selected studies target diverse user groups. Most interactive studies are targeted towards expert (AI and domain) users while many studies published under collaborative and human-centered AI do not specifically identify the target user group.}
  \label{fig:user_groups1}
  % \Description{Sankay chart showing the distribution of studies for user groups.}
\end{figure}

Participants' expertise is a concrete indicator of interactivity as it directly shows the stake of tasks involved.
For example, a complex medical application would require a skilled domain expert for effective system use.
Alternatively, a chatbot can have less knowledgeable users to test its efficacy and prowess. 
The boundary to classify the targeted users intermixes often with varying tasks/needs, for instance, by combining user groups based on the proximity of their usage personas, e.g., experts and domain users put in the same group or domain experts and novices being in the same group. 
Personas are an effective tool in human-AI interaction to identify and differentiate between user groups and their needs (e.g., abilities and preferences) to build unique experiences complementing stakeholders' contexts \cite{holzinger2022personas}.
Personas are designed for specific user groups and their needs to influence interactions with systems.
Figure \ref{fig:user_groups1} shows a distribution of studies included in this survey according to reported user groups.
Across categories, 29 (53\%) studies provide generic adaptations of AI interactions for all user groups.
This also includes the studies where specifics of target users are not reported. 
Excluding the studies with all types of users, 14 studies are targeted at domain-expert users, while only 6 are specific to AI/ML experts. 

Evaluating these patterns, we can assume that user testing is essential for effective user integration with AI systems.
We observe that user testing with actual (targeted) users is not widely practiced and often only involves AI experts \cite{miller2017explainable, koch2017design, kaluarachchi2021review}.
End-user involvement mediated by expert users does not always portray how actual users might behave with the AI system.
This could be acceptable for an XAI system alluded to facilitate AI experts but not for most stakeholders (end-users) \cite{xu2023transitioning, amershi2014power}. 
Domain experts or novices often lack AI expertise and may require more interpretations. 
For instance, Kuang et al. \cite{kuang2023collaboration} and Ottoboni et al. \cite{ottoboni2022multifunctional} show that domain experts are novices to AI while using it for their professional work.
Sometimes, it is necessary to understand how the AI works before putting it into practice for task facilitation.
Some work has focused on educating solely the novice users who lack both domain and AI knowledge \cite{yang2020hybrid, agostinelli2021designing, andersen2022integrating}.
For example, Mishra and Rzeszotarski \cite{mishra2021designing} teach non-experts about AI and enable them to reuse it for their work.
However, studies largely focus on assisting users in their tasks interactively without making them understand how the AI works. 

Oppositely, user exclusion is a significant concern as it reduces human control over systems.
Nakao et al. \cite{nakao2022toward} highlight user exclusion might lead to monopolies of AI systems. 
They urge giving control back to users to bring fairness and agency over AI for decision-making. 
Balancing user control and system autonomy is pivotal, as human-in-the-loop learning is as good as the end user input \cite{chignell2021human}. 
Yet, with knowledgeable users, the degree of control can be enhanced to balance how the systems affect the users \cite{ramos2020interactive}.
A more practical example is depicted in a novice-user educational case study, Teachable Machine \cite{carney2020teachable}, which allows users to build and learn AI experimentally.
But as the domain gets complex, or end-users become less expert in AI, the degree of interactivity gets lower. 
For instance, Cai et al. \cite{cai2019human} showed that providing more control to pathologists directly impacts acceptance, debating for more accessible AI systems for end-users. 
Feng and McDonald \cite{feng2023addressing} underline this sentiment in finding the challenges for practitioners while designing ML applications.
It is also an issue for domain-expert users, as they lack significant knowledge about changing AI paradigms and developments. 

\textbf{\textit{Summary.}} We see numerous studies fail to properly include/test end-users with AI systems. 
Some studies counteract it by including AI professionals as end-users, shifting the goal to largely achieving the task rather than making users aware of AI work.
However, this limits the agency and control of the end-users who are being affected by the AI system in action. 
Recently, studies have emphasized user inclusion in the loop weighing the benefits of making AI more accessible.
Yet, these studies (HCAI and collaborative) are often too generic for user inclusiveness and spread out for diverse and targeted user groups. 
Despite some evidence of empowering users, the overall trend shows resistance toward making AI more open to end-users.

\subsection{AI Implementations}
Systems with concrete implementation details affect the end user behavior and can be evaluated through the type of study, i.e., how it is designed to show interactivity (i.e., the solution), the application area it targets, and types of data it handles during the interaction.
The study type usually but not always affects other parameters such as interface types, learning methods, user goals, and experiences.   
The following subsections summarize the themes and directions in literature targeted at different AI implementations and data modalities that affect interactivity with systems.

\subsubsection{Implementation of the Solution}
\begin{figure}[h]
  \centering
  \includegraphics[height=5cm]{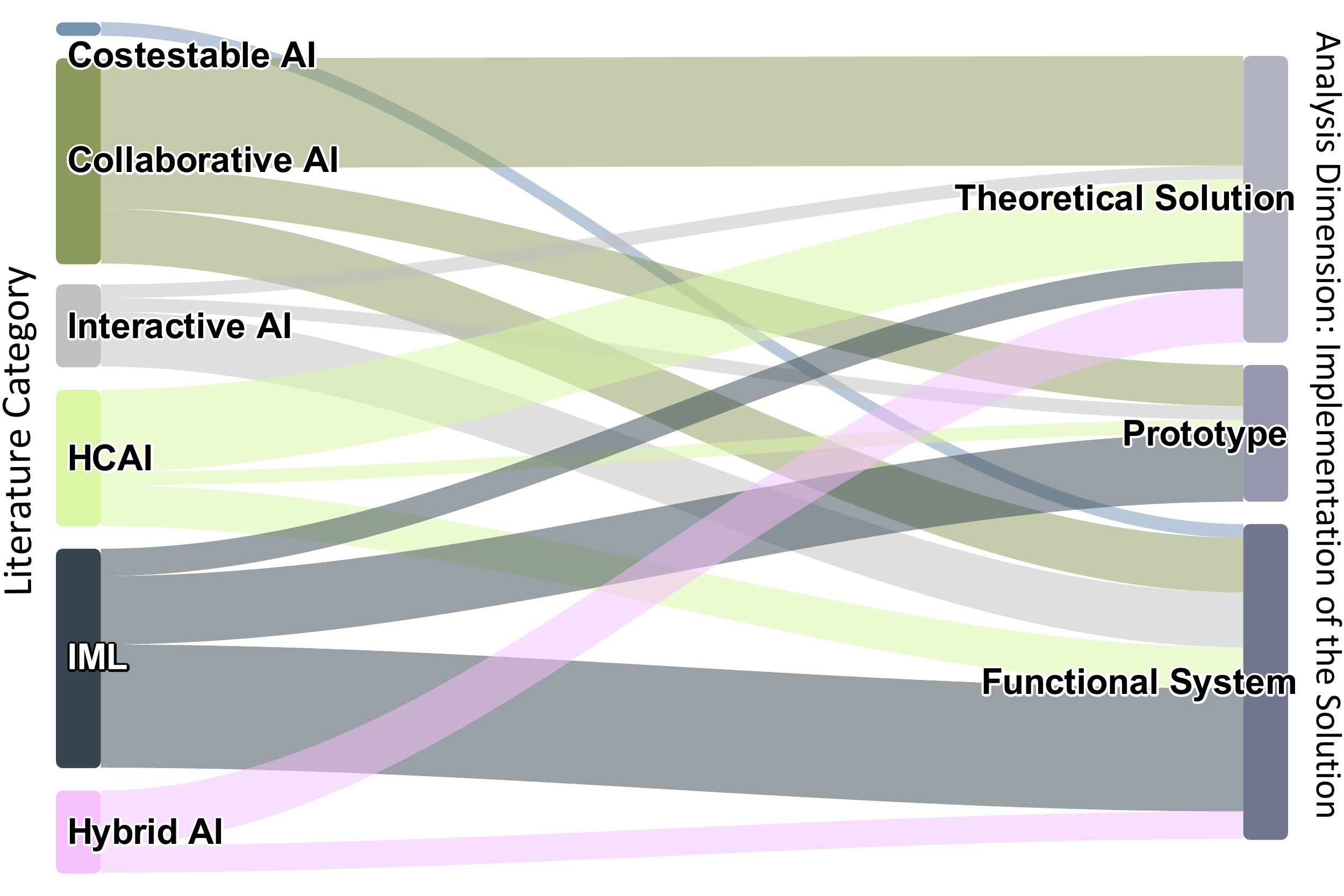}
  \caption{A major portion of collaborative, human-centered, and hybrid AI studies still lies in theoretical proposals of intended systems. Interactive implementations (IML, IAI) are more developed in terms of the practical implementations of their solutions.}
  \label{fig:solution2}
  % \Description{Sankay chart showing the number of studies with the implementation level.}
\end{figure}

The level of (proposed) implementation is proportional to interactivity in the system.
The objective of evaluating the implementation of solutions provides key insights to differentiate practical AI systems for specific contexts and proposals for general acceptability. 
For this purpose, we include studies presenting an actual working system, theoretical solution, or proposal (e.g., prototype) for user interactions in AI/ML systems.
Figure \ref{fig:solution2} provides an overview and distribution of studies according to the level of the active solution implemented. 
We observed that 23 (43\%) studies present some form of system implementation or model for direct interaction with users \cite{wexler2019if, fails2003interactive, carney2020teachable, patel20193d, liu2019mappa}. 
While some studies do not explicitly define a system to interact directly with users but rather implicitly assume interactions in theoretical implementation \cite{fugener2021will, dotov2020dynamic, amershi2014power, shneiderman2020human_rel, teso2019explanatory} or provide proposals such as creating prototypes for testing out hypothesis \cite{kuang2023collaboration, agostinelli2021designing, mruthyunjaya2020human, mishra2021designing, heer2019agency, nakao2022toward, ramos2020interactive, raees2023four}.
Interactive AI/ML applications substantiate their work by building prototypes or interactive systems. 
However, many studies that advocate for collaborations still lag in actual system implementation. 

Influential works, such as those by Shneiderman \cite{shneiderman2020human, shneiderman2020human_rel}, highlight idealized frameworks having high levels of human control and computer automation.
The main direction of his work make use of human knowledge without negating autonomy.
Such solutions propose an idealized human-AI collaboration, empowering people and promoting trustworthiness.
Likewise, a pure collaboration would rely heavily on users to teach the learners and correct their decisions. 
However, overly relying on users to improve learning can also backfire as it can limit the trust of the users in the system's capabilities  \cite{honeycutt2020soliciting}.
Hence a more balanced or hybrid framework is desirable e.g., to balance the AI computational limitations through human intelligence \cite{morteza2019collective}.
However, these all are protean concepts when it comes to human-AI interaction that require a substantial implementation to achieve it, which is lacking.
For instance, Huang and Rust \cite{huang2022framework} hypothesize classifying AI based on the type of functions it performs and focus on high-level AI tasks.
Their work draws aspirations from the biological spectrum, while the current AI is data-focused.
Therefore, idealized goals are not translated into real systems in most cases. 

With more practical systems, collaborative expeditions have shown promising results.
Agostinelli et al. \cite{agostinelli2021designing} and  Mruthyunjaya and Jankowski \cite{mruthyunjaya2020human} provide practical examples of achieving human-AI interaction through collaboration and achieving better user satisfaction. 
Modes of interaction also play a role in defining the level of interaction.
For instance, Kuang et al. \cite{kuang2023collaboration} show that users ask more questions through text-based input than voice-based input.
However, the implementation of this work is still performed through a Wizard-of-Oz method to impersonate an AI agent.
Alternatively, Lee, Liang, and Yang \cite{lee2022coauthor} examined the role of textual input in examining Language Models to inform interaction design.
In terms of implementation, Google’s Teachable Machine \cite{carney2020teachable, withgoogle2023} and What-If \cite{paircode2023, wexler2019if} are far more developed interactive ML/AI tools practically empowering users to interact with AI systems, test their solutions, and use them subsequently.
In terms of other applications, a few examples explored virtual reality interactions, allowing users to create designs with interactive human-AI agents \cite{ruttkay2006human, urban2021designing}.
Linking interactivity with XAI, Teso, and Kersting \cite{teso2019explanatory} highlight explanatory interactions, allowing users to constantly provide feedback on those and explain the reasons behind making a decision.

\subsubsection{Application Domain}
\begin{figure}[h]
  \centering
  \includegraphics[height=5cm]{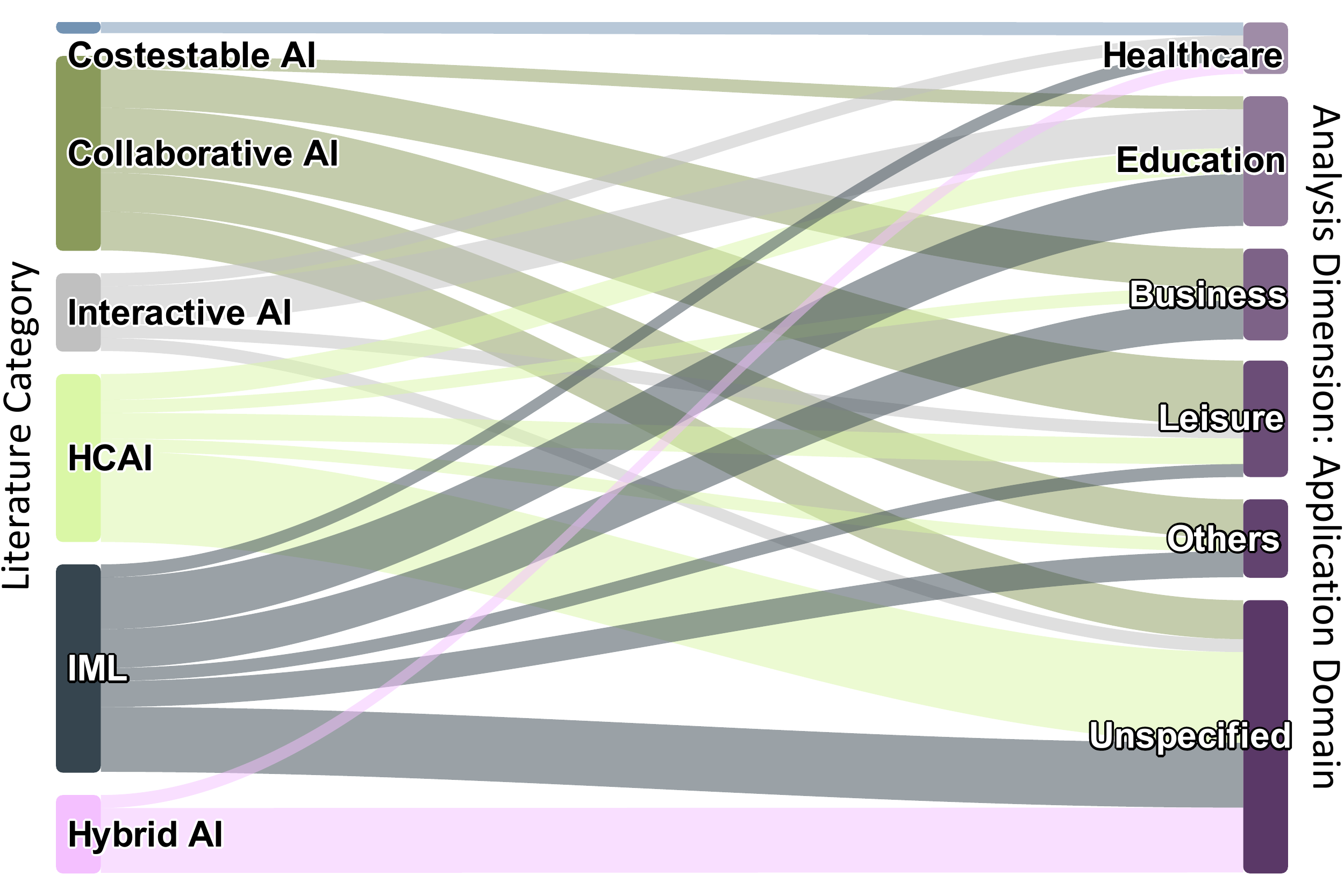}
  \caption{A substantial number of studies do not specify the target application domain (lacking implementation also does not convey much about the application domain). Implementations are also scarce in high-stakes domains such as healthcare, with most implemented solutions targeting education, leisure, and other low-stake domains.}
  \label{fig:app_domain1}
  % \Description{Sankay chart showing the prevalence of application domains.}
\end{figure}

User inclusion for target domains can provide indications and trends of efforts in the type of tasks explored.
Interactive AI application areas are very diverse, hence, we combine those into major categories such as health, education, business, and leisure. 
Figure \ref{fig:app_domain1} shows a collective overview of broader application domains, with the majority being in educational or leisure categories.
Around 33\% of the studies show no specificity in applications, coincidentally, around the same percentage of studies have theoretical solutions. 
Observing the themes in the application domain, we postulate that human interactivity is more tolerated in some areas than others.
For instance, playing around an insignificant decision (e.g., leisure activity) is not equivalent to a consequential one (e.g., medicine, fairness).
This signifies limitations in the capacity of current interactivity in handling uncertainties that users can bring in high-stake tasks.
In our evaluation, only a handful of studies (4 out of 54) are applied in a high-impact application area (e.g., medicine). 
Out of the studies explicitly providing the target application domain, around 33\% (10 out of 30) were found in education/training while around 23\% (7 out of 30) in leisure activities (sports, arts, design), respectively corroborating the tolerance effect.  

We see the main motives of the applications are establishing interaction and providing open-ended suggestions to users to make informed decisions \cite{chignell2021human, mruthyunjaya2020human, doncieux2022human, sanfeliu2020collaborative, kambhampati2020challenges}.
Looking into high-stake applications such as health, Cai et al. \cite{cai2019human} focused on empowering pathologists to search existing images, a common practice in medicine, to aid in the diagnosis of diseases. 
Ottoboni et al. \cite{ottoboni2022multifunctional} defined a more interactive approach to investigating patients with brain injuries. 
Despite being applications of the same field, one focuses on improving diagnostic performance while the other focuses on interactivity to improve the quality of life for both patient and caregiver.
Chiang et al. \cite{chiang2023two} tested their work in a high-stake domain (fairness), but users were recruited from online platforms, not simulating the real-world scenario, which is a widespread concern.
Their work compares groups and individuals in human-AI collaborative recidivism risk assessment.
They conclude that groups relying on AI models are more confident when overturning incorrect AI recommendations, making fairer decisions, and giving AI more credit for correct decisions.
Overall, the high-stake domains are highly under-explored in interactive AI endeavors.

\subsubsection{Modalities of Data}
\begin{figure}[h]
  \centering
  \includegraphics[height=5cm]{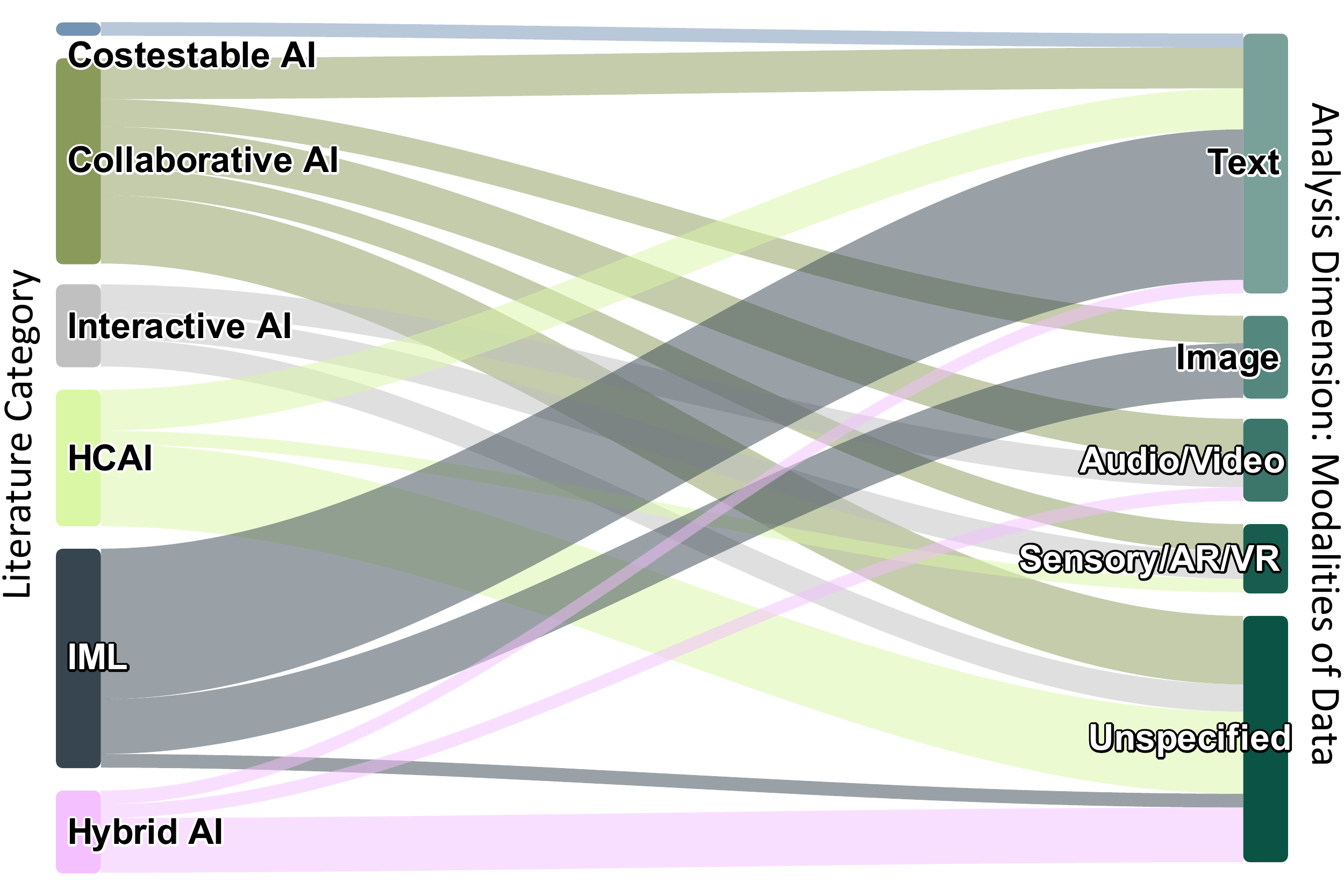}
  \caption{Lacking implementation, many studies also lack insight into the data being handled across categories except interactive machine learning. Text data is prevalent across categories, with many applications targeted at conversational systems.}
  \label{fig:data_modal1}
  % \Description{Sanky chart showing the data types for each category.}
\end{figure}

Data modalities impact interaction types (e.g., prompt vs graphical) and user satisfaction with AI systems. 
For this purpose, we analyzed studies based on the inclusion of data types to evaluate their characteristics.
Following data classifications from the literature, we categorized studies based on data modalities (e.g., text, image, audio/video, and AR/VR/sensory) as shown in figure \ref{fig:data_modal1}.
Textual data is a common interaction mode in the majority of studies (around 33\%) for tasks such as in natural language processing systems.
This is prevalent across AI categories including collaborative or interactive works. 
Image and audio/video (time-based sequential data) data are also commonly used in signal processing or vision systems with the main theme of recognition and classification. 
Some studies employ a combination of these data types or embed other types as well. 

Combined with the applications, textual data is commonly used in conversational interfaces such as in chatbots e.g., in educational/training settings \cite{agostinelli2021designing, andersen2022integrating, maiden2023designing}.
These interactions mainly facilitate student learning through exploration rather than directly providing solutions to the questions. 
Conversational approaches are also seen in robotic applications such as in voice-based interactive educational robots \cite{yang2020hybrid} designed to perform human-centered interactions \cite{mruthyunjaya2020human, pandey2018pepper, patel20193d, sanfeliu2020collaborative, ruttkay2006human}.
Some of these explorations require gestures and AR/VR systems. 
For example, such interactions support building systems for the musicians facilitating them in creating and writing their music through human-AI collaboration \cite{d2015heroic, nicholls2018collaborative}.
Rising complexity, Cabitza et al. \cite{cabitza2023rams} investigated human-AI collaboration in medical diagnosis, and their participants provided diagnoses to the system in a written format. 
The positive effects of combining various data modalities are highlighted in the literature. 
For instance, Kuang et al. \cite{kuang2023collaboration} compared the impact of textual and audio prompts with AI assistants.
Explanatory debugging by Kulesza et al. \cite{kulesza2015principles} is another example of exploring approaches with data to show users how the system makes the decision, and subsequently, users correct the system to improve its learning process. 
We observed that interactive machine-learning studies, being more practical, provide more specifics on the type of data being handled.
Fails and Olsen \cite{fails2003interactive} set the foundational work of interactive machine learning, highlighting its dire need for end-users.
With collaborative AI expeditions, there are more opportunities to experiment with different data modalities together. 
For instance, immersive experiences explored by Xie et al. \cite{xie2019vrgym} and Urban Davis et al. \cite{urban2021designing} tackle complex forms of data modalities. 
However, there are still limited explorations of multi-modal data types other than text and image data for different applications.
 
\textbf{\textit{Summary.}} Interactive AI has a range of theoretical to practical implementations in various low-to-high-stake application areas. 
Most collaborative and hybrid AI applications are generic, focusing on broader perspectives of interaction and AI integration in society rather than implementing solutions to test.
Practical solutions mainly focus either on interactive teaching of the underlying algorithm or improving user performance.
Wherever the interactivity is put into focus, studies aim to empower users in the decision-making process facilitating their tasks.
Likewise, diversity can also be observed in the application domains. 
However, recent research focused more on the low-risk areas (education, leisure, sports, etc.) than the higher-risk areas (healthcare, fairness, security, etc.). 
In addition, many interactive AI studies lack an explicit explanation of the types of data being handled/used. 
Most studies are under-explored from the multi-modal data perspective which is a fundamental challenge in practical user contexts.
Considering the lack of concrete implementations, there lies a potential to grow interactive AI towards more practical experimentation in high-risk areas. 
Across domains, HCAI and IML offer more possibilities for concrete interactions. 
Within IML, several approaches include the user in the machine learning process and various methods (conversational, immersive, and parameter-based) focus on enhancing AI performance in the user domain.

\subsection{AI Goals}
AI studies are targeted at diverse baseline and secondary goals.
To contextualize AI systems' goals to our work, we consider user inclusion as a way forward for being \textit{``human-centered''}. 
Goals either involve users to improve the user experience or allow active modifications to the system. 
On broader levels, goals range from explainability to interactivity. 
Goals can be identified from various types of AI functions such as providing predictions, statistics, or guiding users. 
In practical studies, the goals are defined and measured quantitatively by specifying details about participants, recruitment methods, and other descriptive factors.
Goal definition also depends upon the target audience and task complexity.
For instance, studies that employ simple tasks that do not require significant domain knowledge often have more users compared to studies with complex and domain-specific tasks, as it gets difficult to acquire many people with domain-specific skills and AI expertise. 
The recruitment methods also vary, as the nature of the task is less complex, more studies use crowdsourcing or other online recruitment methods to test their applications.
However, not all studies in our analysis include users directly for testing their solutions, which is a substantial barrier to making interactivity prevalent.
The following sections will dig more into the goals of AI systems ranging from improving user experience to user-led system modifications.

\subsubsection{User Experience}
\begin{figure}[h]
  \centering
  \includegraphics[height=5cm]{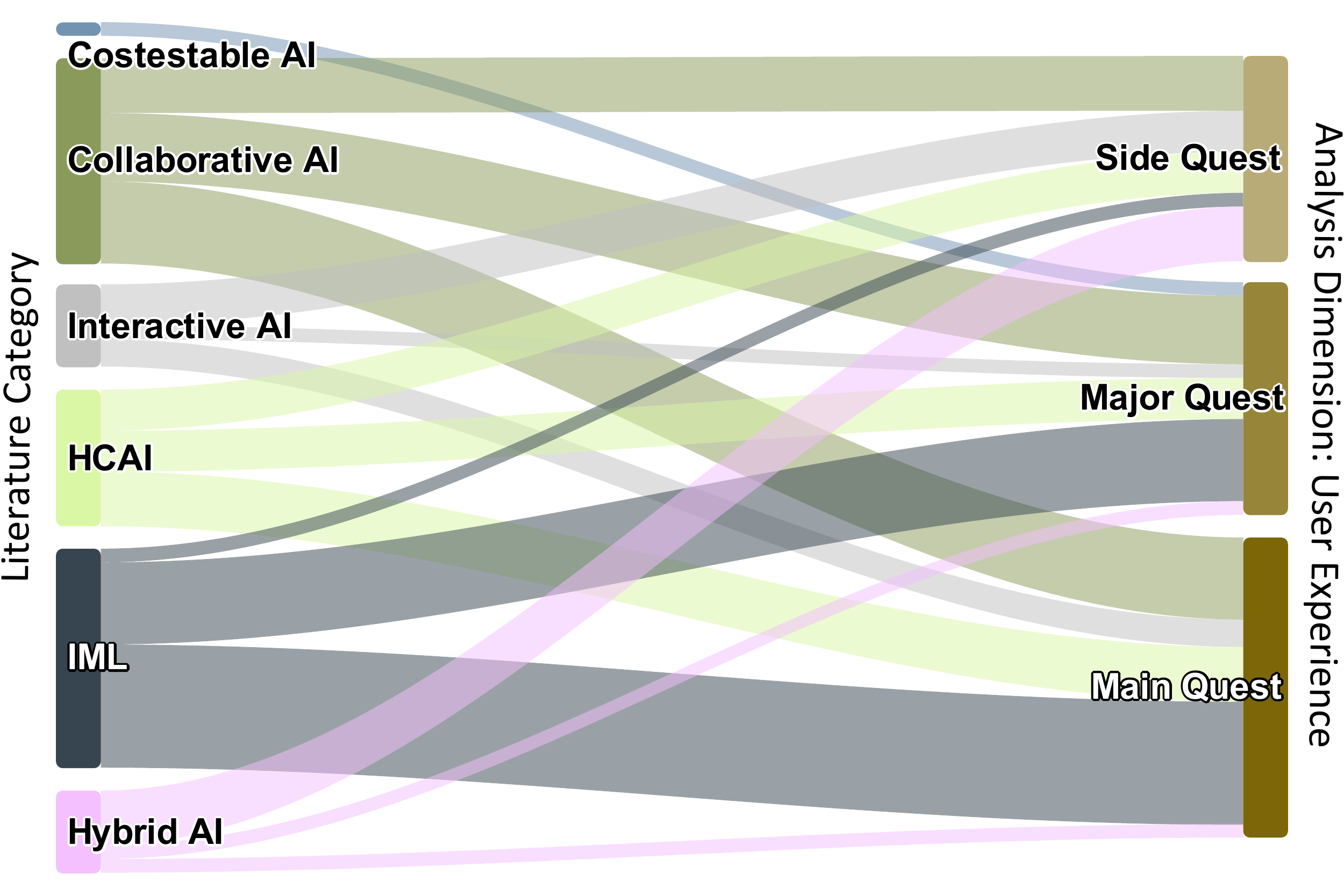}
  \caption{User experience is central to Human-centered studies across categories. A large proportion of studies tackle user experience as their main or major quest.}
  \label{fig:user_exp2}
  % \Description{Sanky chart showing the importance of UX.}
\end{figure}

Despite not all papers tackling user testing or experience in their research, user interaction is a central goal. 
However, some studies focus primarily on enhancing user experience with specific or generic AI systems, for instance, to evaluate the user behavior with the system or to provide better interaction against a defined set of usability guidelines, heuristics, and models.
Figure \ref{fig:user_exp2} shows the goals of the systems in terms of focusing on the user experience.
Comprehensively, 39 (72\%) studies (implicitly) advocate for improving user experience with AI systems with underlying goals.
The majority of the studies treat user experience as an essential element for system acceptability. 
As we highlighted earlier, users are the core of the human-AI loop that enhances interaction \cite{shneiderman2020human_rel}.
Fundamentally, this concept revolves around enhancing collaboration and striving for a more supportive and tool-like architecture.
Despite the obvious benefits, improving user experience comes secondary to creating a collaborative experience firsthand for both human and AI agents.
For example, Stefik \cite{stefik2023roots} discussed cognitive roots and requirements for collaborative AI, going through rigorous evaluation to build effective collaboration. 
This work highlights principles for generating smooth experiences between AI and humans by minimizing mental gaps. 
Some studies \cite{xu2023transitioning, patel20193d} provide guidelines on improving user interaction, highlighting underlying goals such as improving user experience.
Guidelines often explicitly focus on improving experience while the user interacts with the system \cite{amershi2019guidelines}.

User experience has been itself in a broad spectrum of HCI studies and is widely studied across technologies.
Improving user experience makes AI systems more accessible to end users and aligns systems with human cognitive abilities. 
Improving user experience with new technologies that encompass AI is even more important.
These technologies can include implementations in different realities, experiences, and interaction methods.
For instance, an immersive application \cite{xie2019vrgym} is a typical example that falls under this umbrella.
Despite focusing on other tasks, user interaction should also improve the users' immersive experiences. 
Likewise, with an increased infusion of AI into different technologies, exploring user experience is becoming more relevant, for instance, to support AI design (e.g., even to understand user needs), or to guide users to achieve their tasks with AI effectively.
Guiding users and providing explanations is an effective guideline for user experience \cite{amershi2019guidelines}.
Work from Teso and Kersting \cite{teso2019explanatory} which primarily focuses on explanatory system modification by users, shows improving user experience can also help users build trust.
The study also shows interactivity with explanations, allowing users to constantly provide feedback and explain the reasons behind making a decision.
Understandably, user experience takes the central goal in such applications where AI is prominently used in technologies in which users have limited experience.
Therefore, improving the understanding of such systems is a major design guideline for user experience \cite{amershi2019guidelines}. 

\subsubsection{Transparency and XAI}
\begin{figure}[h]
  \centering
  \includegraphics[height=5cm]{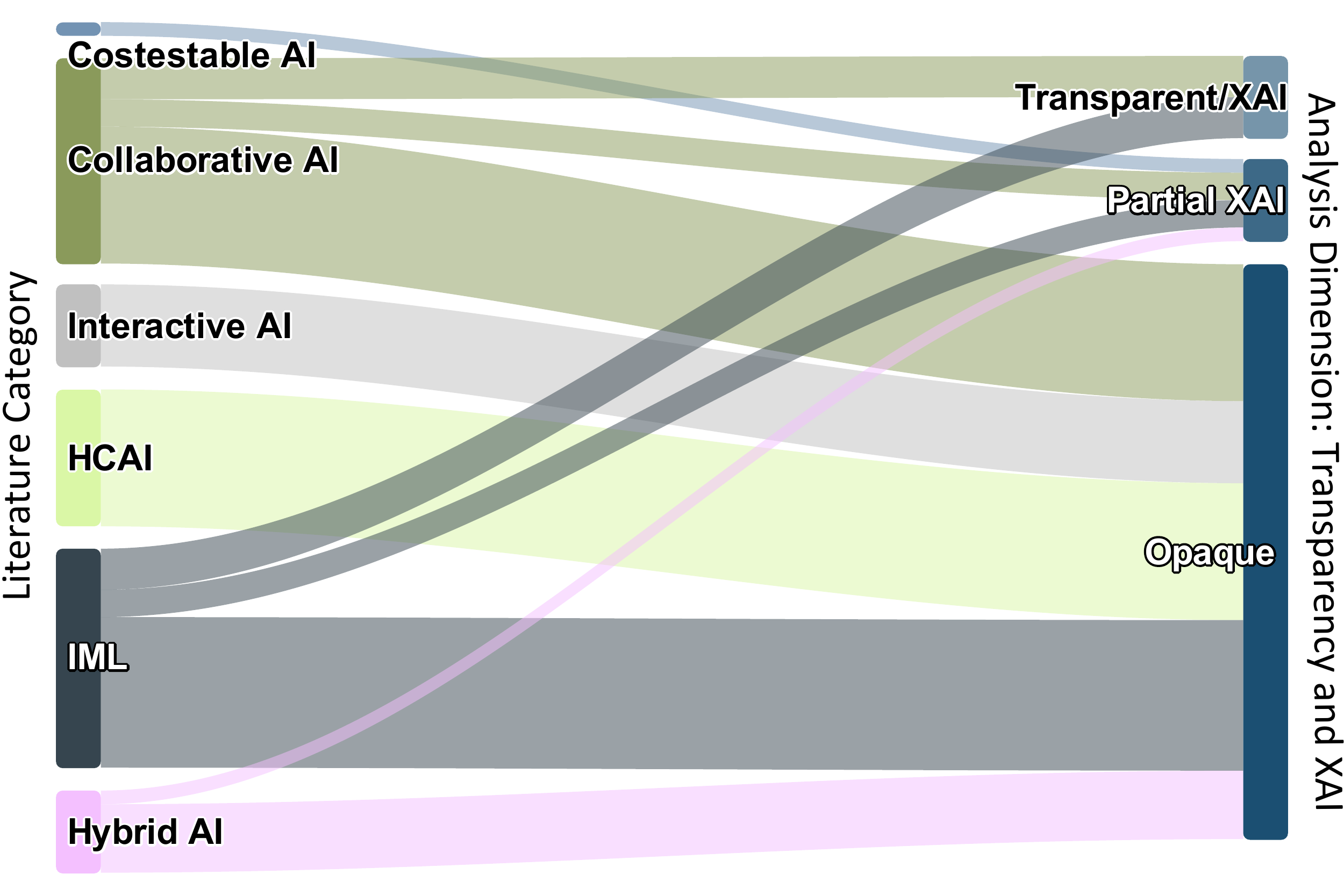}
  \caption{Despite the calls for transparency and explainability for AI models. Opaque AI is widespread across all categories, including interactive and collaborative AI. Only a small number of studies on interactive machine learning and collaborative AI provide explainability built into interactions.}
  \label{fig:XAI4}
  % \Description{Sankay chart showing opaque studies spread across.}
\end{figure}

Transparency is the ability of an AI system to be apparent and traceable in its functionality. 
Transparency underscores the system's adoption and usage across many domains.
However, many current implementations of AI systems are opaque to end-users. 
XAI efforts have increased the explainability of new implementations, however, many still fail to replicate transparency in the end-user domain.
XAI is central to supporting trust, acceptance, and satisfaction with the end users that directly impact the system adoption (e.g., by helping users understand system mechanics) \cite{wang2019designing, amershi2014power}.
Understanding mechanics facilitates the user's mental model by alleviating algorithmic complexities. 
Oppositely, lack of transparency and control decreases trust and user satisfaction with the end-users \cite{cai2019human, botti2006dark}.
Figure \ref{fig:XAI4} shows the prevalence of opaque AI applications spread across all domains calling for a need for more transparent methods.
Our analysis shows that around 11\% of the studies seem to work with transparent AI, while the majority of the remaining studies do not specify the transparency of systems.
We saw considerable research into XAI, but in practice mainly systems contained black-box implementations.
Recently, various applications of more transparent and interpretable methods have been discussed \cite{agostinelli2021designing, wang2019designing, kulesza2015principles, amershi2014power}. 
However, the implementation methods are still lagging in revealing the benefits of openness and increasing the transparency of AI systems. 

Some studies \cite{koch2018group, agostinelli2021designing, lipton2018mythos} consider transparency as an inherent goal to achieve user trust and acceptance.
Still, these are very limited explorations across the vast majority of opaque applications.
As our background work shows, perturbing input to observe how the internal system works is a common method in explaining AI behavior, which itself is not generalizable for different user groups (e.g., novices or domain experts) \cite{wexler2019if}. 
Some applications specifically target transparency to showcase its benefits.
For example, explanatory tools provided by Wexler et al. \cite{wexler2019if} and Cai et al. \cite{cai2019human} have a central goal of discovering the workings of algorithms without providing an explicit explanation.
Teso and Kersting \cite{teso2019explanatory} highlighted that greater interactivity is one key approach being explored to promote transparency and human oversight over AI systems.
Patel and Bhalodiya \cite{patel20193d} sufficed to build explanation specifications but failed to provide an interaction method such as using a natural language to experiment with the system.

Regardless of the goal, transparency needs highly vary from one use case to another (e..g, varying stakeholders' contexts).
For instance, expert doctors often make a medical diagnosis without explicit rules, utilizing knowledge from implicitly gained experience.
The implicit experience would be beneficial for making inferences, yet there would not be a need for over-explaining minor observations during the decision-making process. 
Likewise, the same approach can be applied to AI to make approximate solutions (or transparency) for complex problems rather than perfect solutions for simple problems. 
However, while applying the same principle, applications may have to decide on defining complexities and simple observations. 
Using simpler models would not require defining processes for transparency, however, those do not perform well and complex models are harder to interpret. 
However, simpler models have more subjective control due to hand-engineered features that may become difficult to interpret \cite{koch2017design}.
Therefore, the goal of transparency can be subjected to the complexity of the problem \cite{holzinger2016towards, gigerenzer2008gut}. 
Also, there could be varying levels of transparency and openness depending on the needs of the users/stakeholders.
For instance, having explanations of system behavior (e.g., how the system works and how the decision is made) necessitates that the stakeholder is made aware of system capabilities and limitations, which, to an extent, could also improve the user trust \cite{ribeiro2016should}. 
We see that interactive machine-learning approaches allow more user integration to improve transparency and trust. 
For instance, by allowing direct interaction with the model, providing feedback, or by changing parameters to observe its performance.
Feedback allows users to interact with the AI systems more openly and resonates closely with user expectations. 

\subsubsection{Interactivity}
\begin{figure}[h]
  \centering
  \includegraphics[height=5cm]{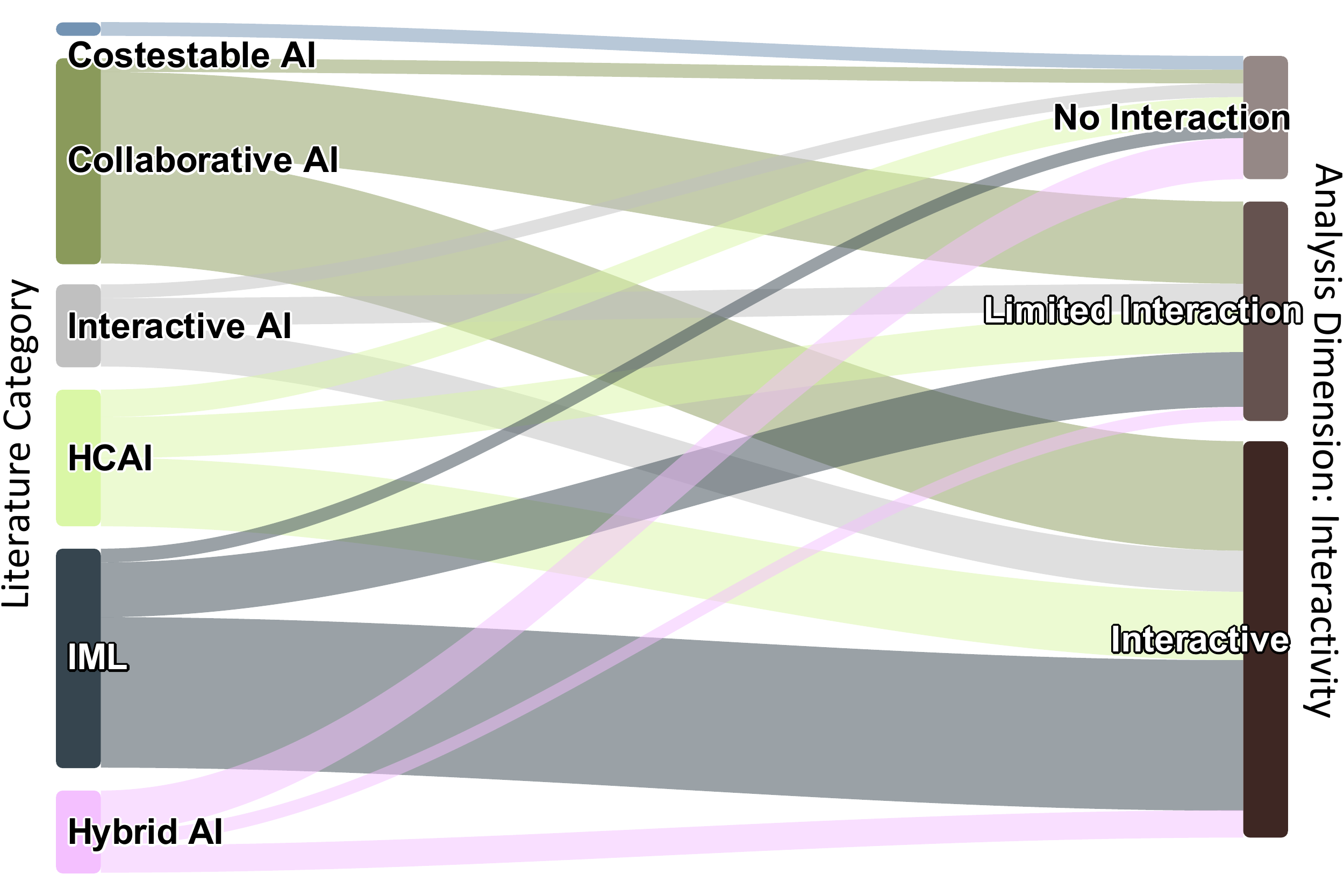}
  \caption{Interactivity is considered central for most of the studies. However, some studies are limited to simple/restricted interactions. Interactivity is prevalent for IML, collaborative, interactive, and human-centered AI.}
  \label{fig:IAI2}
  % \Description{Sankay chart showing interactivity studies across all categories.}
\end{figure}

With the increased penetration of AI into user domains, our focus expands beyond explainability. 
However, limited studies exploit the true interactivity that considers users beyond explanations and experience with AI systems. 
Figure \ref{fig:IAI2} shows interactivity is considered to be part of most AI systems when it comes to allowing users to interact with the system.
Many studies specifically focus on increasing the interactivity for task completion and automating interactions. 
However, the number of studies that allow interactivity to interact with underlying systems is very limited. 
The main principle that interactivity tries to alleviate is the active integration of humans and AI to form teams complementing each other for joint-action tasks \cite{shneiderman2020human}.
These concepts are earlier explored through a perspective of hybrid intelligence or collaborative AI \cite{kamar2016directions, morteza2019collective, peeters2021hybrid, wellsandt2022hybrid, dellermann2019hybrid} to combine efforts of both humans and computers for better collaboration.
Collaborative foundations equip AI systems with access to human intelligence and explore their reasoning capabilities. 
In terms of practical works, most studies \cite{fugener2021will, maiden2023designing, cabitza2023rams, agostinelli2021designing, feng2023addressing} designed assistive tools to automate user tasks. 
For instance, Chiang et al. \cite{chiang2023two} used a conversational collaborative tool to make decisions in criminal applications. 
They show that users who interact more with AI are likely to make good decisions, use their intelligence, and give credit to AI where needed. 
A much more interactive example is created by Maiden et al. \cite{maiden2023designing} to provide solutions to user-generated prompts.
However, beyond interactivity, these studies show limited functionality for users except for assisting in doing some tasks. 

Expanding on the same horizon, interactive AI's primary objective is to maintain user interactions with the system. 
The secondary goal make users part of the process, thereby, improving and adapting AI systems.
Such examples are seen in the interactive ML domain where users lead the AI systems and make corrections when necessary \cite{fails2003interactive, ramos2020interactive, carney2020teachable, raees2023four}.
The interactivity allows rapid feedback, parameter-based adaptation, and correction of underlying models.
For instance, a viable tool developed by Google, Teachable Machine \cite{carney2020teachable, withgoogle2023}, empowers users to take interactive control of AI system development.
Beyond mere interactions with AI systems, only a handful of studies \cite{huang2022framework, mruthyunjaya2020human, holzinger2016towards} allowed interactive adaptations underscoring the true essence of human-AI collaboration.
Practically, Holzinger et al. \cite{holzinger2016towards} allowed a more direct interaction for users, for instance, in a case, to adapt the Ant Colony Algorithm \cite{colorni1991distributed} to use a desirable path (e.g., deemed by the user) in an attempt to solve the Traveling Salesperson Problem \cite{macgregor1996human}. 
Alternatively, an indirect approach by Mruthyunjaya and Jankowski \cite{mruthyunjaya2020human} considers interactivity with users when the AI is uncertain about its decision.
These approaches report the positive influence of including the user in the process and allowing them to make alterations. 
However, the prevalence and complexity of the domains these studies handle are not very diverse. 
To be a more active part of the loop, a user needs to interact with the AI mechanics.

\subsubsection{Algorithm Modification}
\begin{figure}[h]
  \centering
  \includegraphics[height=5cm]{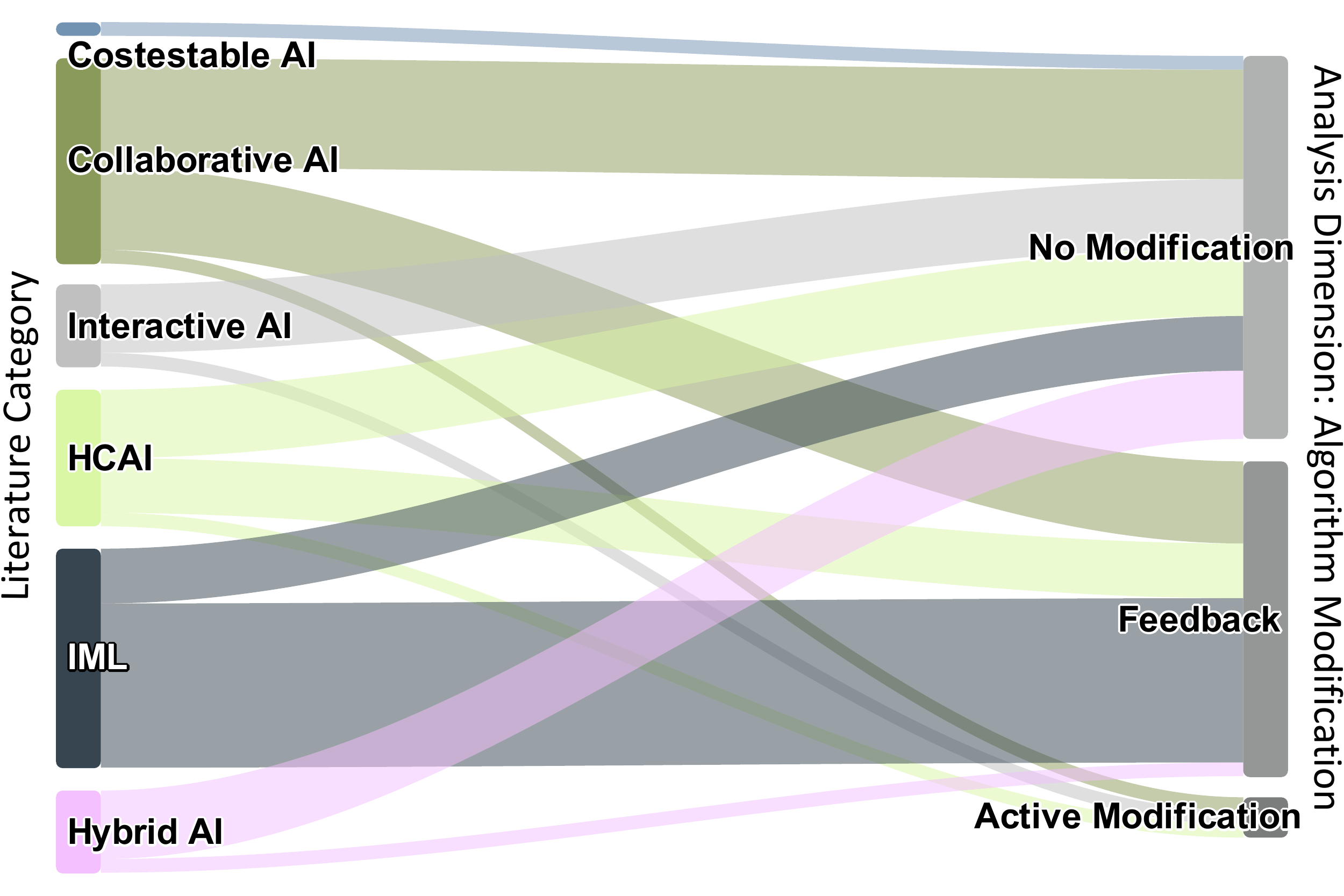}
  \caption{Interactivity has become central to practical AI implementations. However, it is mostly restrained to either interactions or feedback, while only a small number of studies allow users to participate actively beyond feedback.}
  \label{fig:modification}
  % \Description{Bar chart for modification of algorithm for }
\end{figure}

A critical goal of the interaction is the integration of AI around humans as highlighted in research \cite{shneiderman2020human_rel, ramos2020interactive}.
However, the literature evaluation shows that there is little research that is user-centered or allows access to the underlying models.  
Figure \ref{fig:modification} shows a distribution of studies concerning the level of active modification with the AI systems.
In this case, most studies lack active user involvement down to the model/algorithm level, with only marginal (5\%) studies allowing active modification to users. 
One of the several factors that inhibit active modification is the lack of (or the level of) practical implementations. 
Other factors include but are not limited to a lack of user expertise, stake in the task involved, implementation hurdles, or human bias \cite{cai2019human, patel20193d, ottoboni2022multifunctional, lee2022coauthor}.
For instance, Cai et al. \cite{cai2019human} highlight subjective interpretation could lead to bias, and making the underlying model adaptable by end users could lead to incorrect decisions.

The maturity of the implementation is also attributed to some form of feedback or (passive) modification to the system.
For instance, most interactive ML works \cite{feng2023addressing, ramos2020interactive, teso2019explanatory, raees2023four} have some form of feedback mechanism and potentially some active modification.
Overall, the majority of IML works allow modification through parameter feedback. 
In terms of actual active algorithm modification, Nakao et al. \cite{nakao2022toward} present a prototype for end-users to fix potential fairness through algorithm adjustments.
Their prototype allows to make feedback to improve the algorithm and make suggestions by changing the attribute weights.
Doncieux et al. \cite{doncieux2022human} also allow active modification through verbal and non-verbal interaction.
In a different approach for IML, users could train the algorithm on a parameter level through Interactive Machine Teaching \cite{ramos2020interactive}, Explanatory IML \cite{teso2019explanatory}, or experiment with fairness \cite{nakao2022toward, chiang2023two}.
Explanatory IML \cite{teso2019explanatory} shows an increase in system learning performance to correct mistakes twice as efficiently compared to users using a traditional learning system.
For interactive VR experiences, Xie et al. \cite{xie2019vrgym} and Urban Davis et al. \cite{urban2021designing}, report positive effects with the ability they provided to their user to alter their surroundings.
There are only a few examples  \cite{liu2019mappa, huang2022framework, doncieux2022human} beyond parameters-based feedback and alterations. 
This is the point where the users actively engage in the process through explicit, intentional, and informed alterations and decisions.

\textbf{\textit{Summary.}} There are diverse goals when it comes to interaction with the AI system.
We classified these goals as improving user experience, transparency, interactivity, and allowing active algorithm modification.
User experience improvement is a central idea in current research with the majority of studies focusing on user acceptance through enhancing the experience of interaction.
Likewise, transparency of the systems also improves user confidence, satisfaction, and trust. 
However, we see that with current implementations, most of the studies present black box systems while only a few were concerned with adapting the AI mechanics, mostly from the IML domain.
Hence, the interaction landscape calls for more user-inclusive, transparent, user-controlled AI assistance.
Even a simple/restricted user interaction is enough for a feeling of agency, but more work on active modification should be explored.
With the growth in interactive AI applications, the possibilities for diverse interaction for users are likely to grow substantially. 
Therefore, the balance between autonomy and user control is critical for futuristic human AI research. 

\section{Discussion and Directions}
\label{section:5discuss}
The literature exploration shows the breadth and depth of human-AI research, particularly focusing on the constraints of current approaches that go beyond explainability.
Our work makes critical contributions guiding researchers to understand the state-of-the-art, find relevant work in human-AI interaction, and pursue further research. 
We detailed the measures and dimensions employed in the literature to support findings empirically.
The main objectives of current research are depicted through enhancing user experience, accommodating non-AI experts, and creating interactive, and approachable mechanisms to reduce barriers between the AI and its users. 
The current state-of-the-art is slowly progressing towards a more user-centered and interactive approach overcoming the traditional challenges of autonomy.
Further, we discuss the current landscape in practical human-AI studies and provide key considerations in the form of questions to encourage researchers to ponder during their research and/or implementation. 

\subsection{Purpose of Interactive AI}
A fundamental goal of interactivity focuses on building AI around users to facilitate their needs and improve automation.
This goal has a profound impact on end-users/stakeholders in the development and integration of systems in any domain.
For end-users, augmentation with the system could be the top priority to collaborate with the AI effectively, while the business stakeholders may value automation to minimize human interruption to complete tasks, causing a paradox \cite{raisch2021artificial}.
The user-centered approach necessitates providing significant control of the process to the user, and the AI does the heavy lifting of automating the process \cite{shneiderman2020human, shneiderman2020human_rel}.
But why does the AI system allow the user to interact with it, why the interaction has become consequential with the growth of the AI industry in human spheres, and what purpose the interaction can serve as compared to complete isolated automation?
In a nutshell, the question boils down to asking, who is the ultimate decision-maker as AI technology advances?
Human-AI interactivity advocates for user empowerment or user-centered design allowing to actively participate and contest in decision-making.
From the development perspective, it allows systems to capture real-time feedback and make adaptations. 
This is merely a starting agenda that the human-AI systems should consider to improve human-AI collaboration and augmentation, and how these affect the interactivity in AI systems. 

\subsubsection{Collaboration}
One purpose of interactivity is to take the benefits of expertise possessed by humans in domain-specific or even general intelligence tasks.
Human-AI collaboration forms interdependent relationships where AI or users complete the task they are good at, and at the same time, complement the other. 
The notion of AI replacing human tasks is rife for many users and contexts.
However, the majority of implementations aim at improving the user tasks rather than doing tasks by themselves.
The interaction is crucial to improve the performance of the AI system, which can in return improve task performance for the user. 
Experience through evaluated studies depicts that collaboration is an effective approach moving forward \cite{kamar2016directions, peeters2021hybrid}.
Secondly, the problems that are caused by the complexities of AI may not be solved by applying more AI. 
For instance, it could lead to more algorithmic bias in already isolated algorithmic experiments.
Rather, the role of AI with users should be more cooperative and collaborative to achieve better performance. 

The objective for AI systems to interact with humans is to gain more knowledge, gather feedback on their performance, and possibly correct wrong decisions \cite{mruthyunjaya2020human, fugener2021will, carney2020teachable}.
However, how AI and users cooperate and what forms of interactions are achievable is a substantial challenge.
With an increased AI embarking on personal and professional lives, the true spirit of providing control to users favors the AI itself to be more acceptable and trustworthy. 
Research into establishing collaboration and cooperation to form psychological trust is crucial.
Communication and interaction are generally considered effective for forming connections between parties.
Hence, the interactive element is vital in gaining more acceptability, trust, and AI adoption in various domains \cite{feng2023addressing, ramos2020interactive, cai2019human}. 
Therefore, rather than pursuing expert-level human performance, collaboration efforts should focus on how a system can support and enhance human activities (e.g., creativity) \cite{mccormack2020design}.
By doing so, both AI and humans can augment each other by improving the capabilities of each other \cite{doncieux2022human, riedl1901human, heer2019agency}.
But, the focus should rely more on human augmentation rather than system dominance. 
Having defined that goal, open questions revolve around identifying the correct context, the level, and the benefits of interactivity as being excessively interactive also inhibits acceptance \cite{honeycutt2020soliciting}.

\subsubsection{Augmentation}
As we stated earlier, the focus of the development may lie in either augmenting the users or the components of the AI systems, for instance, to improve the system (through users) or facilitate the users (through AI).
Facilitating users to achieve their tasks i.e., augmenting the users to improve task performance, is one of the main AI goals \cite{marshall2022human, cai2019human, huang2022framework, strauch2017ironies}. 
The ideal objective of interactivity, at least with current advances, should be on alleviating human difficulties in achieving tasks through AI, thereby, making it more like any other computer tool \cite{shneiderman2020human}.
In its essence, AI is not like any other tools we had before, therefore, studies should focus on a practical human-AI loop.
Balancing the AI and human needs is essential to remove unwanted human feedback and perceptions \cite{honeycutt2020soliciting}, for instance, through formal specifications to avoid human input errors to the best extent.
Augmentative goals can also be identified from the theoretical solutions to put more effort into implementation to identify user intent.
Users want more control, requiring more fine-grained interactive specification tools, cooperating in their preference to be the initiator of critical actions and avoid interruptions as they formulate their intent \cite{heer2019agency}.

To reduce human errors, attempts can be made to minimize their unwanted interference in the AI loop \cite{xu2023transitioning}.
For instance,  Mruthyunjaya and Jankowski \cite{mruthyunjaya2020human} use a strongly system-centered approach, only allowing humans to intervene when according to the AI, it does not have confidence in its decision. 
Still, even narrowly, the augmentation relies on the human operator, and deciding that balance is more crucial than ever before \cite{huang2022framework}. 
It is also essential for an AI system to augment itself through users.
Without the feedback, it is challenging to improve the efficacy of complex AI systems in the user domain. 
For instance, if the AI system does not augment human tasks, how would it be able to learn from it?
Even if the training has taken place, do we consider that human knowledge has come to stand still or is it a continuous phenomenon?
This leads to fundamental questions that are beyond the context of this study about the evolution of human intelligence to create new knowledge. 

\subsection{Interactive AI in Practice}
While the AI hegemony persists, human-centered approaches also need to adapt to map diverse and new endeavors of interactions.
Research has focused on user involvement abstractly for general acceptability and adaptability with humans in utopian or futuristic scenarios. 
The current landscape, still minimally, is directed towards human-centered AI by advocating for more open and transparent approaches.
However, with widespread societal implications, interaction is the ultimate solution and explanations are merely a starting point.
Even where the user aspects are considered, most work relies on the evaluation of the system curated by AI experts.
The actual users are mostly limited to providing feedback to the developed system with little degree of control.
Mostly, the feedback does not affect the AI system but rather is stored somewhere for the information of AI experts for improving the system in the next iteration. 
Interaction design should incorporate feedback to allow user-centered system improvement. 

\subsubsection{Human-AI Guidelines}
Recent approaches advocate for building AI from a user perspective.
For instance, various studies and tech giants have outlined guidelines for building AI systems that value user needs \cite{amershi2019guidelines, wohlin2014guidelines, googleaiguidebook}.
Valuing AI needs necessitates identifying user needs first, yet most studies emphasize harnessing data rather than user expertise \cite{sandkuhl2019putting}.
In practice, interactive AI has shown a potential to alleviate acceptance barriers and augment users/AI systems \cite{teso2019explanatory, agostinelli2021designing}.
Therefore, it is essential to understand who the user is and whether it is beneficial for both the user and the AI to form an interaction.
Not all users would require the interaction to be of importance, and not all the users would be useful to be part of the interactivity due to the complexity they bring \cite{chignell2021human, kamar2016directions}. 
An important question for designers is to conduct user evaluation for forming effective interactions with AI by defining/following guidelines.
Analysis of users' knowledge about AI is also critical for effective interaction. 
Our analysis corroborates that domain experts are more beneficial to be included in the loop as compared to AI experts and novices for system improvement \cite{miller2017explainable}.
However, our inclination towards this claim could be due to the prevalence of studies targeted to fit the needs of specific tasks for domain experts.
Hence, there is a constant need to explore the role of the users with specific/generic AI applications. 
Interactive ML applications have progressed more in contextual settings, allowing users to make direct interactions with systems.
These implementations are merely existential tools that support user tasks with ML integration underpinning broader objectives of AI. 
This is how we see the current AI research progressing, and perhaps, in practice, using AI may become any other automation tool in the future. 
Still, the broad questions remain unanswered about making AI (even as a tool) more acceptable and controllable by the users. 

\subsubsection{Technical Challenges}
Another important question is to examine the technologies and tools available to allow interaction in real implementations.
Whether the current implementation methods capable of handling interactions at scale the complexity of the models have gone?
Our analysis showed a wide range of implementation endeavors, as some are theoretical solutions and do not cater to the technicalities required to make the interaction possible. 
We observe that there is a trade-off between the stake of the task and the interactivity level (i.e., the high-stake tasks have low interactions and vice versa). 
There are technical challenges currently that inhibit the direct adoption of AI systems to user contexts \cite{wellsandt2022hybrid}. 
Research in interactive AI requires multi-disciplinary concepts and technicalities to be stitched together to form a cohesive user experience. 
Such interactions are hard to achieve and pose challenges to achieving more human-centered implementations.
Most applications are found in assistive tasks such as training/assisting, and with the widespread emergence of language models, a large section of work boiled down to interactive AI assistants.
The general AI implementation follows complex training and optimizing cycles over a large amount of data.
It is inquisitive to investigate how to cater to the interactive integration of users into deployed AI systems and how technologies can support quick model adaptations with obvious challenges.

\subsection{AI and User Agency}
Interactivity is essential for users to exercise agency with the AI systems in an otherwise stale influence. 
AI studies have been predominantly focused on autonomous control of systems for improved efficiency and decision-making. 
Human input is considered to improve AI systems through experience enhancement, task automation, feedback, and prompt correction, overall improving user engagement. 
However, the prevalence of such agencies is minimal, as most traditional AI research focuses on building large and complex systems in isolation and having autonomous control. 
Additionally, the complexity also creates barriers for user agency as complex models are difficult to interact with.
As compared to applications of the recent past, trends show higher interaction, having extended agency for users to interact with the AI system, its output, and the means to reach the goals.
With increasing possibilities of practical implementations and the benefits of building AI through interactive ways, for instance, IML studies are slightly influencing the user agency for algorithm modification.
User agency, contextually, means that the user is actively allowed to modify the mechanics of the AI system, which is often lacking.
Human agency is desired in wide contexts, especially in high-stake domains such as health, legal, privacy, etc. 
However, as the domain gets complex, studies rely more on automation than human agency, which is not ideal for futuristic human-AI interaction research. 
Enhancing agency is pivotal for the critical evaluation of AI systems by experts in the field. 
We see this as a way forward for the enhancement of the interaction research where more interactive systems are built, allowing higher user agency.
However, there are still many factors to consider for optimizing the balance between autonomy and user control. 

Research does demonstrate the benefits of active user alterations to algorithms \cite{marks2007finding, puntoni2021consumers, dietvorst2018overcoming}.
As an example, Auernhammer \cite{auernhammer2020human} discusses the humanistic design research for future AI in a theoretical manner.
Even if it is being perceived in true essence, some solutions are envisioned in more than theoretical proposals.
Offering users the possibility to correct an algorithm’s output, even if only slightly, could be enough to increase the likelihood of using the possibly imperfect algorithm. 
We could consider simply updating a parameter or correcting an output enough for the average end user. 
However, it is essential to investigate whether active modification of the algorithm is necessary.
Also, does providing feedback suffice the purpose in most cases, or is it imperative for users to have control over the interactive experience?
In the practical implementations, we see current practices at least allow users to interact and provide feedback even after the system is deployed. 
Again, a recent trend corroborates user inclusion in AI development and evolution, yet access to the underlying models is still a far-fetched goal.

\subsection{Interfacing AI}
Considering the similar directions of solutions and implementations, studies explored interactivity through various forms of interfaces such as textual (conversational), graphical (manipulation, creation), sensory (robotic), and AR/VR interactions.
The application interfaces in AI can be diverse ranging from simple text prompts to automated AR/VR experiences requiring more practical contexts.
Combining multiple types of interfaces to enhance user interactions and provide better experiences is also a choice. 
Active interaction with AI systems is still an under-explored area of research except for a few tools in the interactive ML domain.
Research on user experience shows that AI implementations are significantly different than the normal web or mobile applications from the past \cite{feng2023addressing}. 
Even if the theoretical solutions are materialized, there is a large gap that does not cover how these systems will be used in practice for (actual) end users.
Interfaces are generally designed to match the goals that users want to achieve.
With new experiences such as AR/VR interfaces, which we categorize as a broad range of interactions captured through one or multiple sensors, user research becomes more imperative than ever. 
For example, using sensors to capture body movements, robot instructions, etc. \cite{xie2019vrgym, mruthyunjaya2020human, doncieux2022human, patel20193d} have different forms of interaction than the usual screen interface. 

Generally, interfaces are categorized into four dimensions, e.g., instructing (teaching), conversing, exploring (evaluating), and manipulating (insights).
Conversing has been an established method of interaction with intelligent systems built upon a perception of communicating with others through conversations \cite{elizaeffect}.
Conversational interfaces give the perception of interactivity through various forms as discussed in our analysis, for instance, text assistants, interfacing with robots, and other sensory inputs.
Conversing is a dominant form of interaction throughout computer and AI research.
It is also prevalent in the current boom of Generative AI systems such as Large Language Models (e.g., ChatGPT \cite{chatgpt}).
For conversational AI assistants, the work of Kuang et al. \cite{kuang2023collaboration} shows an abstract direction to be goal/task oriented from a user perception lens. 
Yang, Oh, and Wang \cite{yang2020hybrid} show an example that practically exhibits interaction with a large-language model. 
Conversational interfaces are powerful tools and can also be used to impersonate experts. 
For instance, marketing companies often employ chatbots to persuade people to think they are interacting with an expert \cite{luo2019frontiers}.
With more technical implementations, we see the benefits of interactions through conversations with AI systems.
However, richer interactions require a shift in conversations from text-based assistants to allow other forms of data. 

Instructing methods are another form of interfacing where users or AI systems take the role of instructor \cite{carney2020teachable, ramos2020interactive}.
Therefore, more usable interfaces like Teachable Machines \cite{carney2020teachable} are highly appreciated and utilized by the users. 
Exploring and manipulating interfaces are comparatively less explored as these are richer forms of interaction and require more implementation effort.
Interfaces allowing interaction through exploring and manipulating are more common in the interactive machine learning domain, allowing richer interactions with the users \cite{kulesza2015principles}.
These interfaces can enable users to explore AI systems about their predictions or manipulate parameters for feedback or alterations.  
With interactive manipulation of the interface objects, users can adapt the parameters of models visually and interact with the underlying models with ease \cite{raees2023four, fails2003interactive}.
For example, to make predictions for the given data and annotate images, subsequently providing details about these actions. 
In addition, a human-AI interface also helps users understand explanations to build causal inferences with the system \cite{holzinger2021toward}.
For instance, interactivity (or feedback), often captured through interfaces, is essential to fathom whether the user can understand AI output.
This also serves as an indicator for continuing the interaction, thereby, building system trust and acceptance.
Understandably, the interaction through some form of interface heavily relies on the level of implementation of the solution. 
Without practical implementations, it would not be feasible to talk about the interfaces and their capabilities.
However, this does not undermine the significance of theoretical contributions as they stand as foundations for experimentation. 

\subsection{Addressing HCAI challenges: current trends and gaps}
The current research in AI strives to address the HCAI challenges by advocating more human involvement in AI development.
The current research emphasizes the significance of transparency and explainability for broader user acceptance of AI systems, for instance, through advancing methods of making models interpretable or focusing on traceable systems.
Some studies transcend beyond mere explanations as a medium of user trust and highlight the benefits of direct and rapid feedback to AI systems, adapting the AI mechanisms, and potentially correcting the AI systems. 
Users add value to the human-AI loop but also bring many uncertainties.
The current research also focuses on improving user interactions with AI systems through various methods to eliminate user errors.
The idealized goal is envisioned through effective collaboration where the user is the ultimate driver stepping out of the shadow of autonomous AI and being the front and center of the human-AI loop.
Newer developments must revolve around humans to be more adaptable and acceptable in personal and professional spheres, complementing human skills and expertise.

To be more human-inclusive, research must mature enough to allow comprehensive human interaction with AI systems. 
This demands user-centric design to investigate how AI systems are designed with human factors in mind from the start.
For instance, AI system feedback and adaptations have proved useful for general acceptability and trust. 
Some studies have focused on the personalizing and customization of AI interactions.
For example, several recent IML studies have shown to alleviate user perception by simplifying the AI systems through user-centered approaches \cite{kuang2023collaboration, ramos2020interactive, raees2023four}.
Xu et al. \cite{xu2023transitioning} call for radical changes in the research agenda for HCAI to make user-centered AI design approaches prevalent and endorsed more often than before.
HCI research objectives should be targeted in the direction of Human-Centered AI to ensure AI systems enhance human capabilities.
The diversity and complexity of AI systems warrant the designers of AI systems to meet users' needs \cite{feng2023addressing}.
AI systems have traditionally focused on specialized audiences (user groups), but with widespread penetration in human life, the psychologies of interaction design need to consider the socio-technical issues of users who are not experts with the AI systems. 
This can include designing interface methods and the creation of natural language-based AI systems that are easy for non-experts.

Inspired by the HCI research, designing interaction is essential for improving user experience with AI systems. 
Our analysis shows that studies have focused on interactivity from various perspectives, such as user experience, task facilitation, or user augmentation. 
Notably, research often does not fit neatly into the defined dimensions and requires approximate mapping to the closest landscapes. 
Interactivity pendulums between users' understanding of the AI to control it, e.g., to contest decisions, provide feedback, or modify it. 
The most common form of interactivity allows users to complete their tasks while at the same time improving the AI system.
We have seen counterexamples where over-interactivity threatens the user's trust in AI systems.
User studies are central to design interactivity to augment the user or the AI system.
However, algorithmic transparency of the AI is another issue in interactions.
The development of AI systems first necessitates addressing issues around transparency and explanations.
Overall, modification of underlying AI models is restrained to educational or experimental applications with low-stakes tasks. 
Additionally, designers need to empower users with the agency to control AI systems to align with their expectations.
The research aspires to move beyond explainability to explore more interactive forms of interaction with AI systems for general acceptance. 
For instance, to design systems that allow user engagement, collaborative decision-making, agency, and adaptations to user needs. 
The need is to improve implementation at a larger level for natural language understanding and integrate explainable AI through interaction.
While not all systems need an explanation, purely deciding to make a system opaque as it is the current state of AI seems to be an odd choice. 
To make progress, radical changes are needed to move towards open systems that are more attuned to user needs and values.
User empowerment and improving the human-AI relationship are expected to be a central theme in future explorations. 

\subsection{Limitations and Future Directions}
Our analysis and evaluation are subjected to some limitations. 
First, the true essence of interactivity lies in \textit{``the involvement of users''}.
However, considering the theoretical solutions in our definition of interactivity includes studies that do not perform actual tests with user involvement. 
Hence, the results of this work should be interpreted with more caution when compared or reported with user-centered studies with participants.
Second, the scale could be a limiting factor as the field of AI is so diverse and changing rapidly.
Hence, extracting only studies related to human-AI interaction in the plethora of information extravaganza is difficult despite employing comprehensive search and evaluation strategies.
It remains plausible that our search strategy might have missed some relevant studies, resulting in insights that may be overlooked or affect this review. 
We acknowledge that restrictive search queries found articles that only use the equivalent language. 
To reduce this effect, we performed extensive snowballing to identify the relevant studies through forward and backward searches.
The search strategy could have been enhanced with more keywords.
On the contrary, most papers showed an overlap in citations, which could indicate a proper saturation of interactive AI literature.
Lastly, this research is also constrained to temporal scope (e.g., up to 2023), and the literature produced after that might affect our synthesis.

Our objective lies in human-AI studies where users are allowed with intentional, explicit, and informed interactions with the AI system. 
Although this is an idealized case, current research merely proposes what is aspired as a human-centered world i.e., humans at the center of the AI loop. 
Technical advances in the interactive machine learning field operationalize this concept where users can make intentional, informed, and explicit interactions with the system.
We argue that allowing balanced control on the adaptation of AI systems to users is essential and could benefit the AI systems in the long run. 
There is a high need to explore user agency to improve interactions with the underlying mechanisms of AI. 
More research efforts can prove the impact of balancing autonomy and user control (agency) in practice.
Collaborative and hybrid approaches are targeted to balance out the best of both worlds by combining intelligence from humans and AI, thereby, reporting user-centered results. 
Hybrid intelligence seems to have a promising outlook on the possibilities of human-centered AI and shared goals surrounding user interactivity. 

Transparency has been a core principle and many studies have been advocating for much-needed traceable and explainable systems.
However, we still see that practical implementations are often opaque. 
Also, the interpretability needs vary depending upon the role of the users with the system. Limitations in practical applications also pose significant challenges for transparency/XAI in identifying suitable AI output evaluation methods. 
Without defined benchmarks for XAI, user evaluation methods are prone to many knowledge gaps depending upon the varying explainability needs of different users.
Despite the arguments that not all AI systems require explanations, questions about the ethical accountability of AI still stand valid.
For instance, establishing frameworks to measure user challenges in anticipating the consequence of interactive AI modifications or tracing back the events leading to the decisions.
Research in high-stake domains such as healthcare, justice, and security are prone to face the aforementioned risks if explanations and transparent design are not built at their core. 
We see more efforts being made for explainability in practical settings, and the field can contribute significantly to the overall adoption of AI.
Future work should continue to strive to loop AI around humans rather than the opposite. 
With the advent of Generative AI, the complexity of models has exploded, making them more complex than ever, thereby requiring more effort for interpretations behind generated content.
Future work could consider the opportunity to include interactivity in Generative AI and explore the possibilities.
With the growing popularity and new creative ways of using AI methods, new possibilities for interactivity seem to arise.
Therefore, future work can also explore the effectiveness of appropriate interaction types for AI systems in different user contexts.

\section{Conclusion}
\label{section:6conclude}
AI systems are currently being applied across domains, resulting in increasingly complex interactions between humans and AI. 
Fundamental concerns arise among research, industrial, and societal stakeholders as a result of these interactions, related to explainability, trust, acceptability, and adoption of AI, among others. 
Despite these concerns, most current literature on human-AI interaction, including surveys, focuses on the topic of explainability, with the area of Explainable AI largely overshadowing other forms of interaction.
In this study, we systematically review the human-AI interaction literature, shifting our focus to studies that extend beyond Explainable AI into the realm of Interactive AI, where user interaction with the AI system is defined as being explicit, intentional, and informed. 
Our analysis identifies key patterns, gaps, and opportunities, across three main dimensions, comprising AI users, implementations, and goals, each containing multiple sub-dimensions. 
Our primary research contribution lies in pinpointing the need for prioritizing the user's role in the AI loop beyond explanations, advocating for active user interaction through feedback and/or co-creation (for example, enabling the users to actively adapt and co-design the AI mechanics).

Overall, our analysis reveals that recent works in human-AI interaction beyond explanations partially consider human influence as part of the loop.
User involvement is not, at the moment, considered an impactful form of interaction, and most research focuses on improving user experience in the interaction with AI systems rather than facilitating more active forms of user interaction and control. 
Indeed, addressing the challenges of inexperienced AI users and creating more approachable AI is highly visible in current research to make systems less intimidating for non-experts. 
Much of contemporary research also focuses on highlighting the need for increased interaction from a theoretical perspective, but practical evidence shows that user-centered approaches are still under-explored.
Our results also highlight a nascent tendency, evident in a handful of studies, to create AI systems that adapt to user adjustments and feedback. 
These studies demonstrate that AI can be used as a tool rather than an autonomous entity.
Yet, due to a lack of practical solutions, this type of research also situates in aspiring for more user interaction, rather than proposing concrete solutions.
Some research tendencies are also noted in the area of interactive machine-learning applications with more specialized (expert) users in highly contextual settings. 
Finally, we find that a large section of the relevant work boils down to interactive AI assistants focusing on teaching or simplifying tasks for humans. 

The issue of insufficient user control and agency remains largely untouched, even in the most interactive studies.
Some studies allow interactive feedback from users to AI systems and vice versa to make users more aware of the internal AI mechanisms. 
However, only a few studies go one step forward by allowing users to modify the algorithm design parameters to practice complete agency over AI decision-making.
Overall, our findings highlight that user agency over AI systems beyond explainability is still very limited, and this is a gap that we expect more studies to focus on in the years to come. 

\section*{Acknowledgments}
We would like to extend our gratitude to our colleagues at Sappi, Markie Janse van Rensburg, and Marjorie Boles for their continued support in the context of the Sappi-RIT Digital Innovation Lab at Golisano College of Computing and Information Sciences, RIT. Any explanations, findings, or conclusions expressed in this work are those of the authors and do not necessarily reflect the views of Sappi. 

 \bibliographystyle{elsarticle-num} 
 \bibliography{cas-refs}





% \end{thebibliography}
\end{document}